# Quantum Study of Dispersion-Corrected Electronic and Optical Properties of syn- and anti- $B_{18}H_{22}$ Clusters with/without Sulfur Doping for Tunable Optoelectronics

Fakher Abbas[1,*], Nabil Joudieh[1,2,3] Habib Abboud[3] and Nidal Chamoun[4,5*]

[1]Department of Physics, Faculty of Sciences, Damascus University, Damascus, Syria
[2]Laser spectroscopy laboratory, Higher Institute for Laser Research and Applications, Damascus University, Damascus, Syria
[3]Faculty of Pharmacy, International University for Science and Technology, Daraa Highway, Ghabagheb, Damascus, Syria
[4]Department of Statistics, Faculty of Sciences, Damascus University, Damascus, Syria
[5]CASP, Antioch Syrian University, Maaret Saidnaya, Damascus, Syria

* Correspondence:
fa858ab95.com@gmail.com, chamoun@uni-bonn.de

**Abstract:**

This study presents a detailed quantum chemical investigation of the electronic and excited-state properties of syn- and anti-isomers of the borane cluster $B_{18}H_{22}$ and their sulfur-doped derivatives.

Using the PB86/def2-SVP level of theory with dispersion corrections, the research examines how sulfur substitution and non-covalent interactions influence cluster stability, electronic structure, spectra, and optoelectronic potential. Comparative analysis of the syn- and anti-isomers reveals the impact of molecular conformation on frontier molecular orbitals and related energetic and photophysical properties. Sulfur doping is shown to enhance charge delocalization, stabilize excited states, and improve thermal stability, which form key factors for tunable laser applications.

The inclusion of advanced dispersion correction methods proves essential for accurately capturing many-body interactions that govern electronic behavior. By elucidating the interplay between structural features, substitution patterns, and computational modeling, the work provides valuable insights into structure-property relationships critical for designing borane-based materials with tailored optoelectronic and thermal characteristics. In addition to doping and dispersion effects, functionals' effects were examined through comparison amidst various ones with different asymptotic exchange on excitation and gap energies.

In summary, the computational methodology combined good DFT/Basis/dispersion set to accurately predict geometries, IR/UV spectra, NMR chemical shifts, dipole moments, polarizability and excited-state properties, thereby offering a comprehensive theoretical understanding of syn-, anti- isomers of $B_{18}H_{22}$ and its sulfur doped variants.

# 1. Introduction and State-of-the-Art Review

Lasers, characterized by coherence, monochromaticity, and directionality are based on stimulated emission in a gain medium under population inversion, producing photons with identical phase, wavelength, and direction. Molecular lasers play vital roles in science /technology, enabling advanced spectroscopic techniques like Raman and laser-induced breakdown spectroscopies [1–6]. Solid (liquid, gaseous) medium lasers like Nd:YAG (dye, CO2) are suited for high-power applications (tunability, industrial cutting) applications, whereas semiconductors are crucial for compact systems. Optimizing the active medium enhances efficiency, stability, and spectral properties [7,8].

Borane-based lasers represent a breakthrough in laser material science. Anti-$B_{18}H_{22}$ clusters exhibit efficient and photostable blue emission at 406 nm with superior quantum yield (9.5%), addressing limitations of organic dyes and quantum dots like low stability. Their structural versatility supports polymer incorporation and potential spectral expansion through chemical modifications .

The experimental/theoretical studies of borane lasers, particularly anti-$B_{18}H_{22}$ and recently the forgotten syn-$B_{18}H_{22}$, is critical for advancing laser material science and addressing the limitations of existing technologies. Boranes, with their unique polyhedral structures and delocalized electronic states, offer exceptional photophysical properties, including efficient blue laser emission, high quantum yields, and superior photostability compared to conventional organic dyes and quantum dots. These attributes make boranes a promising alternative for applications in spectroscopy, materials processing, optoelectronics, and medical technologies [9,10].

Theoretical investigations enable a deeper understanding of the photo physics and photochemistry underlying borane-based lasers. By modeling electronic transitions, energy states, and substitution effects, researchers can optimize borane derivatives for enhanced efficiency, tunability and stability. Such studies also provide insights into how molecular modifications influence absorption and emission properties, paving the way for the development of boranes emitting light across different spectral regions. Incorporating boranes into polymer matrices or nanocomposites further broadens their potential applications in electrooptical devices and luminescent solar concentrators. Ultimately, the study of borane lasers not only introduces a novel family of laser materials but also establishes a foundation for sustainable, cost-effective solutions in cutting-edge laser technologies [11].

In 2014, Cerdán et al. [12] discovered that the borane anti-$B_{18}H_{22}$ was the first boron hybride to act as a laser material, emitting stable blue light (406 nm, 9.5% efficiency). In 2018, Londesborough et al. reacted anti-$B_{18}H_{22}$ with pyridine, forming new clusters—most notably $B_{16}H_{18}$-3',8'-Py (main product), $B_{18}H_{20}$-6',9'-Py (minor product), and $B_{18}H_{20}$-8'-Py (trace)— identified by X-ray and NMR. Among these, $B_{18}H_{20}$-6',9'-Py showed thermochromic fluorescence shifting from red (620 nm, room temp.) to yellow (585 nm, 8 K), while another cluster exhibited weak phosphorescence and the third was non-luminescent. The study highlighted the impact of cluster architecture and ligand dynamics on photophysics and solubility.

In 2019 [13], Londesborough et al. reacted the laser-active borane anti-$B_{18}H_{22}$ with iodine, yielding mono- and di-iodinated derivatives in high yields, displaying strong green phosphorescence ($\lambda_{max}$ = 525 nm, $\Phi_L$ = 0.41; $\lambda_{max}$ = 545 nm, $\Phi_L$ = 0.71) and could generate singlet oxygen with high quantum yields ($\Phi\Delta$ = 0.52 and 0.36), making them promising photosensitizers. In a separate 2019 study, Zhang et al. [14] found that the anionic surfactant SDS significantly enhanced the fluorescence of n-$B_{18}H_{22}$, with the effect related to SDS concentration and its hydrophobic shielding effect. In 2020, methylation of anti-$B_{18}H_{22}$ by Londesborough et al. [15] led to a highly substituted borane with expanded polyhedral volume and improved absorption and solubility, highlighting new possibilities for borane-based laser materials.

Recent studies on anti-$B_{18}H_{22}$ and its derivatives have expanded their structural and functional diversity. In 2020, Anderson et al.[16] synthesized a brominated derivative (4-Br-anti-$B_{18}H_{21}$) with dual emission (fluorescence at 410 nm and phosphorescence at 503 nm), showing potential as a ratiometric oxygen sensor. Chen et al.[17] created an isoquinoline hybrid ($B_{18}H_{20}(NC_9H_7)_2$) with aggregation-induced emission (AIE), significantly increasing quantum yield in water-rich media. Bould et al. [18] developed highly stable, blue-emitting alkylated derivatives (Rx-anti-$B_{18}H_{22-x}$), achieving near-unity quantum yields and highlighting the impact of substitution on photophysics. In 2021, Ševčík et al.[19] examined anti-$B_{18}H_{22}$ and its tetraethyl derivative in solution and polymer matrices, noting high quantum yields in solution but reduced efficiency in solid films, with photostability and solvent effects analyzed via photoluminescence and NMR.

In 2022, Bould et al.[20] synthesized novel Selena boranes by inserting selenium into both syn- and anti-$B_{18}H_{22}$, yielding macro-polyhedral anions such as [$SeB_{18}H_{19}$]⁻ and [$Se_2B_{18}H_{19}$]⁻, characterized by NMR, mass spectrometry, and X-ray crystallography. These clusters displayed unique architectures, and DFT calculations supported their structural analysis. Anderson et al. [21] benchmarked the luminescence and UV stability of anti-$B_{18}H_{22}$ and its halogenated derivatives, finding excellent stability in powders but degradation in polymer films, and demonstrated their utility in prototype UV imaging. In 2023, Ehn et al. [22] synthesized and structurally characterized a range of chlorinated anti-$B_{18}H_{22}$ derivatives, with dichlorinated isomers showing stable blue emission and structure-dependent photophysics.

In all these studies, anti-$B_{18}H_{22}$ was shown to be advantageous for photophysical properties and laser applications, while the syn-isomer has historically been overlooked as nonluminescent.

As a matter of fact, Patel et al. [23] revisited in 2023 the syn-$B_{18}H_{22}$ isomer and demonstrated that crystalline syn-$B_{18}H_{22}$ displays blue fluorescence, with phosphorescence emerging upon specific position substitution. This significant study synthesized and characterized three monothiol-substituted isomers: [1-HS-syn-$B_{18}H_{21}$], [3-HS-syn-$B_{18}H_{21}$], and [4-HS-syn-$B_{18}H_{21}$], identifying two polymorphic forms for isomers with substitution sites at 1 and 4, using several techniques for characterization, such as NMR, mass spectrometry, X-ray diffraction, IR, and luminescence spectroscopy. The study also reported the first mechanochromic luminescence shift in a borane cluster and suggested potential for syn-$B_{18}H_{22}$ derivatives as carbon-free self-assembled monolayer components.

In 2025, Deeb et al. [24] conducted a comprehensive quantum chemical investigation of mono-halogenated derivatives of the anti-$B_{18}H_{22}$ borane cluster, assessing their potential for optoelectronic applications. The research employed (time-dependent) density functional theory, (TF)DFT, using the PBE0/def2-SVPD and B3LYP/6-311+G(d) methods within the ORCA software. The study focused on the parent anti-$B_{18}H_{22}$ and its halogen-substituted variants: 7-F-anti-$B_{18}H_{21}$, 4-F-anti-$B_{18}H_{21}$ and the experimentally synthesized 4-Br-anti-$B_{18}H_{21}$.

In line with [23], we limited the substitution in Boran to Sulfur suspecting a tangible role for the latter in laser optoelectronics. Actually, Sulfur, with variable oxidation states (–2 to +6), is essential for tuning electronic, optical, and charge transport properties, as it enhances singlet-triplet energy gaps, spin–orbit coupling, and charge injection, improving delayed fluorescence and phosphorescence in devices like OLEDs. Moreover, Sulfur doping modifies bandgaps and photoluminescence in materials such as ZnO and $MoS_2$, while its high molar refraction increases polymer refractive indices, and hence Sulfur-rich compounds can act as photocatalysts and boost device efficiency.

Building on this, we investigated in our work the impact of elemental sulfur doping on the "forgotten" borane laser compound, focusing on how sulfur incorporation influences structure, energetics, spectral features, and photophysical properties. This study can pave the way for more complex sulfur-based substitutions and provide the rational design for advanced borane-based optoelectronic materials.

Actually, for the forgotten isomer, only the study in [23] addressed its theoretical characterization, and performed geometry optimization (NMR shielding) using Head-Gordon's long-range-corrected wB97XD functional -with empirical dispersion- and the def2-TZVPP basis set (GIAO method at the same level), while dipole moments and excited-state properties were analyzed using TD-DFT.

Our contribution here is to enrich the theoretical study of [23], showing the effects of using a new method/basis/dispersion correction on the structure and optoelectronic properties and presenting new calculated properties. A comparative study between the two isomers syn and anti is moreover presented in order to understand the role and effects of the geometric structure on the electric, photophysical properties, and the lasing power.

In all, this work reports the first observation of a mechanochromic luminescence shift in a borane cluster, enriching the field of boron-based luminescent materials. Additionally, it provides, in line with [23], an evidence that substituted syn-$B_{18}H_{22}$ derivatives could serve as components in carbon-free self-assembled monolayers, expanding their potential applications.

The plan of the paper is as follows. In section 2, we restate the goals and objectives of the work, while we present the adopted methodology and methods in section 3. In section 4, we detail the obtained results, whereas we end up with conclusions in section 5. Some technical details are presented in appendices.

## 2. Goals/Objectives and Significance

This study presents a comprehensive quantum chemical study focused on the electronic and excited-state properties of syn and anti-isomers of the borane cluster $B_{18}H_{22}$, along with their sulfur-doped derivatives.

Using PB86/def2-SVP level of theory enhanced by dispersion corrections, the research investigates how sulfur substitution and non-covalent interactions influence cluster stability, electronic structure, NMR/IR/UV-V spectra, absorption intensity, static polarizability and optoelectronic potential. The lifetime of excited states is also calculated using the Strickler-Berg relationship.

The study compares the electronic properties of the syn and anti-isomers in order to understand the impact of molecular conformation on frontier molecular orbitals and their influence on energetic and photophysical properties. It then explores the effects of sulfur doping, which is known to improve charge delocalization, stabilize excited states, and enhance thermal stability, which are all key factors for applications such as tunable lasers.

The inclusion of sophisticated dispersion correction methods is emphasized as essential for accurately modeling many-body interactions that significantly affect the clusters' electronic behavior. By analyzing how structural factors -like substitution position and conformation, together with computational approaches- affect electronic and spectroscopic properties, the work provides valuable insights into the structure-property relationships governing optoelectronic performance.

The work findings lay the groundwork for reasonable designs of borane-based materials with tailored electronic and thermal characteristics. Furthermore, excited-state calculations and planned experimental validation may fully characterize the photophysical properties and practical applicability of sulfur-doped borane clusters.

Overall, our study bridges fundamental borane chemistry with advanced optoelectronic technology, highlighting the critical roles of sulfur doping and dispersion corrections in developing next-generation laser materials.

We shall use the Orca6 software package to perform the required calculations and then simulate the results of the output files in special programs in order to draw theoretical photoelectron, NMR, absorption, and fluorescence spectra.

Our findings demonstrate that incorporating sulfur into borane clusters effectively stabilizes excited states vital for population inversion, a key process in laser operation. This doping also modifies the frontier molecular orbitals and enhances the overall stability of the clusters. Sulfur-doped borane clusters show strong absorption and emission across a broad spectrum, from ultraviolet to near-infrared, making them well-suited for laser applications. These improvements establish sulfur-doped systems as promising candidates for tunable lasers and other optoelectronic devices.

Moreover, sulfur integration not only fine-tunes the electronic and spectroscopic properties, but also supports scalable synthesis methods, enabling further molecular substitutions and improving the thermal and photophysical stability of boron-based laser materials.

By bridging fundamental borane chemistry with practical optoelectronic needs, sulfur doping provides a versatile approach for designing advanced laser materials with customized optical performance and enhanced durability. Continued experimental and theoretical work is essential to fully unlock the potential of sulfur-doped borane clusters in future optoelectronic technologies.

## 3. Computational methodology/methods:

The calculations, including geometry optimizations and others, were carried out using the energy gradient that was constructed analytically and implemented in the ORCA 6.0.0 program package [25].

We used the DFT method with the functional PB86, and basis set Def2-SVP. We either dropped the dispersion corrections (D0), or included them (D2 or D3 or D4) [26].

The molecular geometry calculation was performed along with the determination of energetic and photovoltaic properties, dipole moment, and calculation of infrared frequencies, as well as IR, NMR, and UV-VIS spectra. We also performed calculations including the calculation of the excited state lifetimes of the forgotten borane molecule and its forgotten analog with the sulfur substituent according to the Strickler-Berg equation [27].

Chemcraft [28], a graphical interface for drawing, was used in conjunction with ChamDraw to obtain visual representations of the five figures and generate the corresponding IR and NMR spectra [29].

The Python programming language was used to plot the absorption and fluorescence spectra of the forgotten borane molecules without and with the sulfur substituent in order to compare the spectra and results.

## 4. Results and Discussion:

We aim here to calculate/evaluate and compare various properties of the unsubstituted and sulfur-substituted borane molecules. This entails determining the electronic energy, optimized geometry, dipole moment, and polarizability, along with several energy and thermodynamic calculations. Furthermore, we determine the vibrational frequencies, infrared spectra, theoretical photoelectron spectra, excited state lifetime spectra in terms of absorption wavelengths, and nuclear magnetic resonance (NMR) spectra of the 'forgotten' borane molecule in order to confirm its lasing action.

## 4.1 Optimized geometry:

We show in Figure 1, the large polyhedral borane clusters with the molecular structure of the isomers, Anti-$B_{18}H_{22}$ and Syn-$B_{18}H_{22}$, which are derived from the fusion of the two $\{B_{10}\}$ units with surrounding H forming subclusters. The difference between the two structures stem from how these subclusters are connected [23]. In the Anti(Syn)-isomer, the bridging hydrogens are located on opposite (same) side of the polyhedral units, so that the two shared units (B(5) B(6) in one subcluster and their primes in the other) are such that B(5)=B(5'), B(6)=B(6') in a (syn), whereas we have the identification reversed in c (anti): B(5)=B(6'), B(6)=B(5').

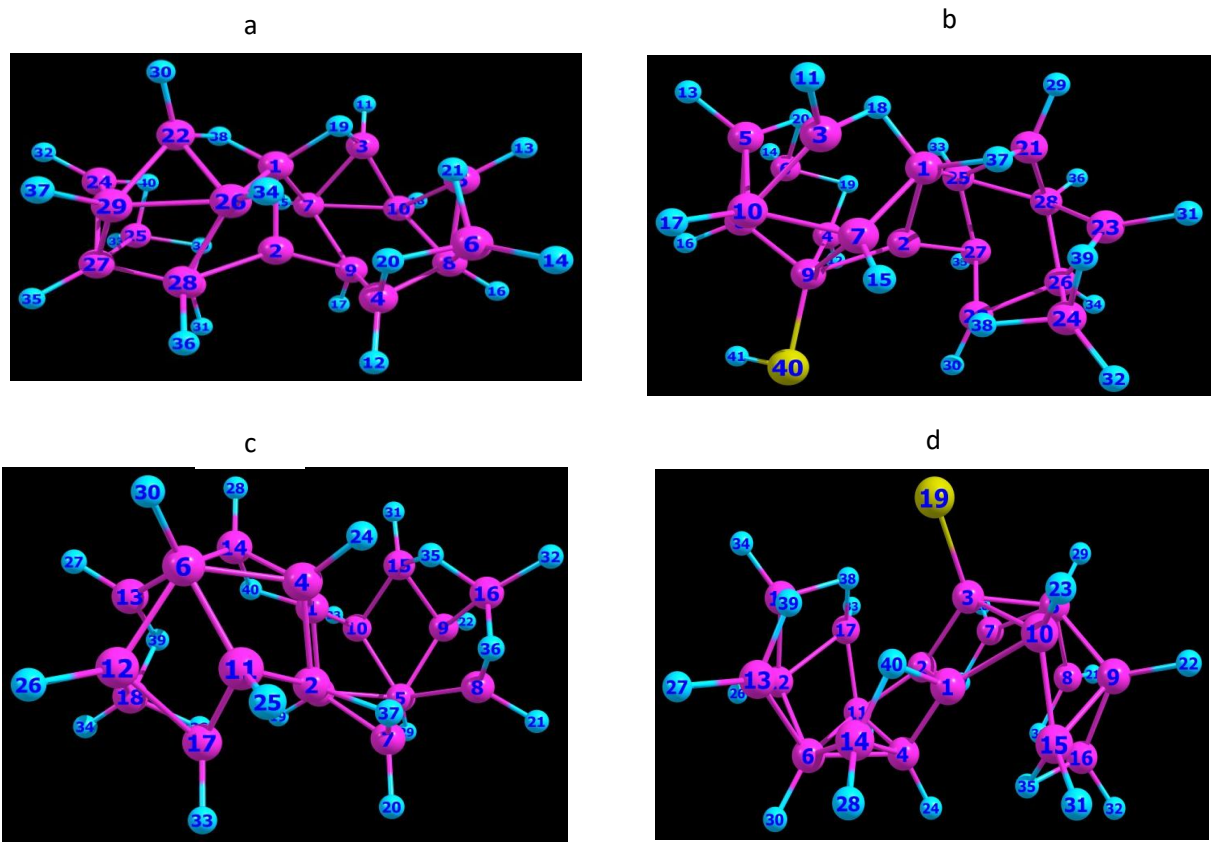

**Fig. 1**. Optimized BP86/def2-svp geometries of: **a)** $B_{18}H_{22} - syn$ Isomer, **c)** $B_{18}H_{22} - anti$ Isomer, **b)** $[1 - HS - syn - B_{18}H_{21}]$ and **d)** $[4 - HS - anti - B_{18}H_{21}]$ isomers. Yellow color denotes a sulfur substituent

This structure gives the syn-molecule an additional mirror symmetry, compared to the anti-isomer, which affects its luminescence characteristics, regarding fluorescence and laser emission. Actually, limited research has been conducted on this syn-isomer, but it has gained attention recently due to its unique crystalline photophysical properties, such as its aggregation-induced fluorescence.

Moreover, the substitutability of sulfur atoms, as it occurs in thiol groups, differs significantly between the two isomers due to variations in their molecular geometries and electron density distributions.

Substitution studies on the anti-$B_{18}H_{22}$ isomer have predominantly focused on synthesizing fluorescent derivatives, with substitution patterns constrained by the compound's inversion symmetry and electron density distribution across its boron vertices. However, systematic investigations into sulfur atom substitution within the $B_{18}H_{22}$ antagonist framework remain underexplored in the literature, particularly when contrasted with analogous research involving other elements, such as halogens, substitution. Computational modeling and experimental analyses have collectively demonstrated that thiol substitution exhibits preferential localization at boron positions B1-B4, a phenomenon attributable to pronounced negative charge density within these regions. Among these four sites, B4 emerges as the thermodynamically favored position for functionalization.

Crucially, thiolation induces a marked disruption of syn-$B_{18}H_{22}$'s double symmetry, precipitating significant geometric reorganization within the cluster architecture. This structural perturbation is empirically validated through observed alterations in interatomic distances and bond angles, as illustrated in Figure 1.d.

**Table 1.** Bond lengths (R in Angstrom), Vibrational frequencies (W in cm$^{-1}$) and angles (A in degrees), for the four system: **a.** $B_{18}H_{22}(anti\ and\ syn"forgetten")$, and **b**. $4HS-B_{18}H_{21}anti\ and\ 1HS-B_{18}H_{21}syn"forgetten")$, at their equilibrium geometries using the DFT- PB86/def2-SVP.

| Bond | $B_{18}H_{22}-anti$ | $B_{18}H_{22}-syn$ | Bond | $4HS-B_{18}H_{21}-anti$ | $1HS-B_{18}H_{21}-syn$ |
|---|---|---|---|---|---|
| R(1-3) | 1.814 | 1.819 | R(1-2) | 1.82 | 1.819 |
| R(3-7) | 1.8 | 1.806 | R(3-10) | 1.761 | 1.803 |
| R(3-19) | 1.206 | 1.322 | R(5-8) | 1.807 | 1.757 |
| R(5-8) | 1.76 | 1.811 | R(5-10) | 1.769 | 1.807 |
| R(5-10) | 1.798 | 1.77 | R(6-14) | 1.206 | 1.768 |
| R(6-14) | 1.766 | 1.206 | A(3-10-5) | 66.7 | 60.3 |
| W(19) | 459.4 | 467.1 | A(5-10-7) | 115.1 | 109.1 |
| W(20) | 509.2 | 511.4 | W(21) | 425.3 | 448.6 |
| W(25) | 570.3 | 575.9 | W(22) | 461.4 | 464.9 |
| W(43) | 714.6 | 711.2 | W(30) | 582 | 592.6 |
| A(5-10-7) | 109.1 | 115.6 | W(31) | 583.3 | 598.4 |
| A(5-8-10) | 107.6 | 59 | W(32) | 595.3 | 601.8 |

A comparative structural analysis of borane isomers, as summarized in Table 1, offers critical insights into the fundamental factors influencing the stability and reactivity of these cluster complexes. The data reveal pronounced structural disparities between the syn- and anti-configurations in both $B_{18}H_{22}$ (a) and its sulfur-substituted derivatives (b), as determined through use of DFT at the PB86/def2-SVP level.

A detailed examination of part (a) of the table highlights distinctive variations in bond lengths that correspond closely with their respective three-dimensional geometries. Notably, the R (3–19) bond length differs markedly between the anti (1.206 Å) and the syn (1.322 Å) isomers, reflecting a redistribution of electron density at these skeletal positions. Conversely, the R (6–14) bond exhibits an inverse trend, measuring 1.766 Å in the anti- relative to 1.206 Å in the syn- form, indicating a substantial structural reorganization accompanying isomerization. These findings underscore the intricate relationship between electronic structure and geometric configuration within these borane clusters.

The angular parameters further underscore significant conformational distinctions between the isomers of part (a), exemplified by the A(5–8–10) angle, which undergoes a pronounced change from 107.6° in the anti-isomer to 59° in the syn configuration. This substantial angular distortion likely reflects a fundamental reorganization of the polyhedral cage architecture induced by the isomerization process.

As to the substitution of hydrogen with sulfhydryl group (HS), shown in Table 1b, It imparts profound electronic and structural perturbations to the borane framework. Comparative analysis of 4HS-$B_{18}H_{21}$-anti and 1HS-$B_{18}H_{21}$-syn isomers reveals that sulfur incorporation significantly modifies the cluster's electronic environment, as evidenced by distinct variations in bond length distributions. Specifically, the R (3–10) bond length differs notably between the 1HS-syn (1.761 Å) and 4HS-anti (1.803 Å) configurations, indicating that the position of sulfur substitution critically influences electron delocalization pathways within the polyhedral skeleton.

Correspondingly, the angular parameter A(3–10–5) measures 66.7° in the 1HS-syn isomer compared to 60.3° in the 4HS-anti counterpart, suggesting that sulfur incorporation induces differential strain patterns across the cluster framework contingent upon both substitution site and isomeric form.

Furthermore, vibrational frequency analyses (denoted by W values) demonstrate that both isomerization and sulfur substitution exert significant effects on the dynamic properties of these clusters. Notably, the W21 vibrational mode shifts from 425.3 cm$^{-1}$ in the 1HS-syn isomer to 448.6 cm$^{-1}$ in the 4HS-anti isomer, indicative of altered force constants associated with specific vibrational motions. Collectively, these findings highlight the intricate interplay between substitution position, geometric configuration, and the resulting electronic and dynamic characteristics of borane clusters. This phenomenon likely underpins the observed differences in thermodynamic stability between the isomeric forms. The variation in the W32 vibrational frequency, 595.3 cm$^{-1}$ for the 1HS-syn isomer compared to 601.8 cm$^{-1}$ for the 4HS-anti-isomer, further substantiates the systematic influence of sulfur incorporation on the vibrational characteristics of these borane clusters.

The structural disparities observed between syn and anti-isomers, coupled with the position-dependent effects of sulfur substitution, elucidate fundamental aspects of electronic reorganization within polyhedral boranes. The introduction of a sulfur heteroatom induces an asymmetric distribution of electron density, likely generating localized electron-rich sites that modulate the cluster's chemical reactivity. Pronounced distortions in bond lengths and angles indicate that sulfur substitution significantly alters the three-dimensional geometry of the borane cage, potentially giving rise to unique reactive centers amenable to catalytic applications or advanced materials design.

Moreover, the electronic perturbations associated with sulfur incorporation are expected to affect the HOMO-LUMO gap, with consequential implications for the optical and electronic properties of these clusters. Collectively, this comprehensive structural and electronic analysis advances our understanding of the interplay between isomerization and heteroatom incorporation in polyhedral boranes, thereby establishing critical structure-property relationships that are foundational for the rational development of boron-based materials with tailored functionalities.

### 4.2 Electronic/electric properties & Atomic charge:

The isomerization dynamics of syn- and anti-$B_{18}H_{22}$ boranes are intrinsically governed by their distinct electronic architectures, which directly influence optoelectronic behavior and charge distribution patterns. DFT enables precise quantification of frontier molecular orbitals (FMOs), dipole moments, and electron density reorganization during structural transitions [23].

The atomic charge distribution in syn- and anti-$B_{18}H_{22}$ isomers, calculated via Löwdin population analysis at the DFT-PB86/def2-SVP level, reveals critical electronic reorganization patterns that govern isomerization energetics and thermodynamic preference. We present the data in Tables 2 and 3.

**Table 2.** Mulliken and Löwdin charge analysis of: **a.** [$B_{18}H_{22} - syn"forgetten"$] **b.** [$B_{18}H_{22} - anti$] isomers at their equilibrium geometries using the DFT method at PB86/def2-SVP basis set.

**a.**

| number Atom | Atom Label | D0 Löwdin charge | D0 Mulliken Charge | D2 Löwdin charge | D2 Mulliken Charge | D3 Löwdin charge | D3 Mulliken Charge | D4 Löwdin charge | D4 Mulliken Charge |
|---|---|---|---|---|---|---|---|---|---|
| 6 | B | -0.11864 | 0.039745 | -0.11816 | 0.020439 | -0.11799 | 0.032332 | -0.11799 | 0.032283 |
| 7 | B | -0.11842 | 0.275989 | -0.11751 | 0.258622 | -0.1176 | 0.267986 | -0.1176 | 0.268111 |
| 8 | B | -0.09456 | 0.198277 | -0.09106 | 0.194463 | -0.09597 | 0.207182 | -0.09597 | 0.207131 |
| 9 | B | -0.10221 | 0.119971 | -0.09857 | 0.119034 | -0.10292 | 0.128276 | -0.10292 | 0.128349 |
| 10 | H | 0.027743 | -0.02558 | 0.027859 | -0.02702 | 0.02707 | -0.02728 | 0.02707 | -0.02727 |
| 11 | H | 0.030468 | -0.02562 | 0.030825 | -0.02985 | 0.029494 | -0.03175 | 0.029494 | -0.03176 |
| 12 | H | 0.027837 | -0.01563 | 0.02776 | -0.01779 | 0.026654 | -0.01958 | 0.026654 | -0.01959 |
| 13 | H | 0.027862 | 0.030174 | 0.02785 | 0.029369 | 0.026908 | 0.031735 | 0.026908 | 0.031736 |
| 14 | H | 0.023348 | -0.06526 | 0.023235 | -0.06403 | 0.023008 | -0.07124 | 0.023008 | -0.07122 |
| 15 | H | 0.028791 | -0.06757 | 0.028285 | -0.06897 | 0.027972 | -0.06841 | 0.027972 | -0.06845 |
| 16 | H | 0.038806 | -0.09195 | 0.0386 | -0.09341 | 0.03812 | -0.10825 | 0.03812 | -0.10826 |
| 17 | H | 0.041291 | -0.06493 | 0.040606 | -0.06494 | 0.040375 | -0.07138 | 0.040375 | -0.0714 |
| 18 | H | 0.170859 | 0.073325 | 0.171634 | 0.071215 | 0.171397 | 0.060738 | 0.171397 | 0.060751 |
| 19 | H | 0.163843 | 0.092821 | 0.164797 | 0.096218 | 0.164177 | 0.090736 | 0.164177 | 0.090741 |
| 20 | H | 0.151886 | 0.0288 | 0.15219 | 0.027344 | 0.152338 | 0.015786 | 0.152338 | 0.015775 |
| 21 | B | -0.08694 | -0.03272 | -0.08677 | -0.02744 | -0.08539 | -0.03232 | -0.08539 | -0.03237 |
| 22 | B | -0.04093 | 0.045561 | -0.04246 | 0.074063 | -0.03944 | 0.078703 | -0.03944 | 0.078801 |
| 23 | B | -0.06716 | -0.14894 | -0.06633 | -0.13855 | -0.06485 | -0.13647 | -0.06485 | -0.13655 |
| 24 | B | -0.0769 | -0.17614 | -0.07923 | -0.18398 | -0.07527 | -0.18416 | -0.07527 | -0.18418 |
| 25 | B | -0.11864 | 0.03973 | -0.11816 | 0.020442 | -0.11799 | 0.032341 | -0.11799 | 0.03231 |
| 26 | B | -0.11842 | 0.275977 | -0.11751 | 0.258644 | -0.1176 | 0.267986 | -0.1176 | 0.268116 |

**b.**

| number Atom | Atom Label | D0 Löwdin charge | D0 Mulliken Charge | D2 Löwdin charge | D2 Mulliken Charge | D3 Löwdin charge | D3 Mulliken Charge | D4 Löwdin charge | D4 Mulliken Charge |
|---|---|---|---|---|---|---|---|---|---|
| 5 | B | -0.0769 | -0.17613 | -0.07923 | -0.18396 | -0.07527 | -0.18417 | -0.07527 | -0.18417 |
| 6 | B | -0.11864 | 0.039745 | -0.11816 | 0.020439 | -0.11799 | 0.032332 | -0.11799 | 0.032283 |
| 7 | B | -0.11842 | 0.275989 | -0.11751 | 0.258622 | -0.1176 | 0.267986 | -0.1176 | 0.268111 |
| 8 | B | -0.09456 | 0.198277 | -0.09106 | 0.194463 | -0.09597 | 0.207182 | -0.09597 | 0.207131 |
| 9 | B | -0.10221 | 0.119971 | -0.09857 | 0.119034 | -0.10292 | 0.128276 | -0.10292 | 0.128349 |
| 10 | H | 0.027743 | -0.02558 | 0.027859 | -0.02702 | 0.02707 | -0.02728 | 0.02707 | -0.02727 |
| 11 | H | 0.030468 | -0.02562 | 0.030825 | -0.02985 | 0.029494 | -0.03175 | 0.029494 | -0.03176 |
| 12 | H | 0.027837 | -0.01563 | 0.02776 | -0.01779 | 0.026654 | -0.01958 | 0.026654 | -0.01959 |
| 13 | H | 0.027862 | 0.030174 | 0.02785 | 0.029369 | 0.026908 | 0.031735 | 0.026908 | 0.031736 |
| 14 | H | 0.023348 | -0.06526 | 0.023235 | -0.06403 | 0.023008 | -0.07124 | 0.023008 | -0.07122 |
| 15 | H | 0.028791 | -0.06757 | 0.028285 | -0.06897 | 0.027972 | -0.06841 | 0.027972 | -0.06845 |
| 16 | H | 0.038806 | -0.09195 | 0.0386 | -0.09341 | 0.03812 | -0.10825 | 0.03812 | -0.10826 |
| 17 | H | 0.041291 | -0.06493 | 0.040606 | -0.06494 | 0.040375 | -0.07138 | 0.040375 | -0.0714 |
| 18 | H | 0.170859 | 0.073325 | 0.171634 | 0.071215 | 0.171397 | 0.060738 | 0.171397 | 0.060751 |
| 19 | H | 0.163843 | 0.092821 | 0.164797 | 0.096218 | 0.164177 | 0.090736 | 0.164177 | 0.090741 |
| 20 | H | 0.151886 | 0.0288 | 0.15219 | 0.027344 | 0.152338 | 0.015786 | 0.152338 | 0.015775 |
| 21 | B | -0.08694 | -0.03272 | -0.08677 | -0.02744 | -0.08539 | -0.03232 | -0.08539 | -0.03237 |
| 22 | B | -0.04093 | 0.045561 | -0.04246 | 0.074063 | -0.03944 | 0.078703 | -0.03944 | 0.078801 |
| 23 | B | -0.06716 | -0.14894 | -0.06633 | -0.13855 | -0.06485 | -0.13647 | -0.06485 | -0.13655 |
| 24 | B | -0.0769 | -0.17614 | -0.07923 | -0.18398 | -0.07527 | -0.18416 | -0.07527 | -0.18418 |
| 25 | B | -0.11864 | 0.03973 | -0.11816 | 0.020442 | -0.11799 | 0.032341 | -0.11799 | 0.03231 |
| 26 | B | -0.11842 | 0.275977 | -0.11751 | 0.258644 | -0.1176 | 0.267986 | -0.1176 | 0.268116 |

The Table 2 data present both Löwdin and Mulliken atomic charges for two borane isomers (syn- in a, and anti- in b), providing nuanced insights into their electronic structures and bonding environments. Although both charge partitioning schemes are derived from the same underlying electronic structure calculations, they differ fundamentally in their mathematical frameworks for distributing electron density among constituent atoms, resulting in notable variations in both the magnitude and, in some cases, the sign of the computed atomic charges.

More precisely, Löwdin charges on boron atoms range narrowly from –0.11864 to –0.04093, whereas Mulliken charges exhibit a broader distribution spanning from 0.27599 to –0.17614. Of particular interest, boron atoms 6, 7, 8, and 26 display positive Mulliken charges despite corresponding negative Löwdin values, underscoring the divergent nature of these population analysis methods.

The incorporation of dispersion corrections (D2, D3, and D4) exerts minimal influence on both Löwdin and Mulliken charge distributions across the isomers, as evidenced by the approximate constancy of charge values irrespective of the applied correction. This observation suggests that dispersion interactions do not substantially perturb the electronic charge distribution within these borane clusters.

The observed differences in charge distribution between Isomers (a) and (b) are anticipated to influence their respective reactivity profiles. For instance, the positive Mulliken charges calculated for boron atoms 6, 7, and 8 in Isomer ($B_{18}H_{22}$-syn) suggest that these positions may exhibit enhanced susceptibility to nucleophilic attack compared to their counterparts in Isomer ($B_{18}H_{22}$-anti).

**Table 3.** Mulliken and Löwdin charge analysis of: **a.** [ $1HS - B_{18}H_{21} - syn"forgetten"$] **b.** [ $4HS - B_{18}H_{21} - anti$] isomers at their equilibrium geometries using the DFT method at PB86/def2-SVP basis set.

**a.**

| number Atom | Atom Label | D0 Löwdin charge | D0 Mulliken Charge | D2 Löwdin charge | D2 Mulliken Charge | D3 Löwdin charge | D3 Mulliken Charge | D4 Löwdin charge | D4 Mulliken Charge |
|---|---|---|---|---|---|---|---|---|---|
| 5 | B | -0.08482 | -0.17801 | -0.08632 | -0.19212 | -0.08503 | -0.18169 | -0.08533 | -0.18216 |
| 6 | B | -0.12914 | 0.055944 | -0.12847 | 0.049112 | -0.12849 | 0.053054 | -0.12844 | 0.054024 |
| 7 | B | -0.13764 | 0.27035 | -0.13706 | 0.247685 | -0.13694 | 0.256582 | -0.13704 | 0.259449 |
| 8 | B | -0.24816 | 0.056928 | -0.24537 | 0.042252 | -0.25365 | 0.040183 | -0.25267 | 0.044548 |
| 9 | B | -0.10492 | 0.179305 | -0.10146 | 0.181898 | -0.10421 | 0.1855 | -0.10427 | 0.184018 |
| 10 | H | 0.028534 | -0.02798 | 0.028724 | -0.02922 | 0.028797 | -0.02892 | 0.028774 | -0.02859 |
| 11 | H | 0.032903 | -0.02868 | 0.033544 | -0.03237 | 0.033802 | -0.03166 | 0.033673 | -0.03104 |
| 12 | H | 0.028004 | -0.02122 | 0.027955 | -0.02337 | 0.028113 | -0.02303 | 0.028069 | -0.02265 |
| 13 | H | 0.028392 | 0.027184 | 0.028414 | 0.026127 | 0.028351 | 0.026363 | 0.028396 | 0.0266 |
| 14 | H | 0.026214 | -0.05658 | 0.026019 | -0.05572 | 0.026697 | -0.05609 | 0.026621 | -0.0561 |
| 15 | H | 0.026968 | -0.07183 | 0.025815 | -0.07561 | 0.026628 | -0.07536 | 0.02665 | -0.07474 |
| 16 | H | 0.044819 | -0.05572 | 0.043872 | -0.05694 | 0.04485 | -0.05726 | 0.044807 | -0.05696 |
| 17 | H | 0.17139 | 0.068741 | 0.172077 | 0.066468 | 0.172298 | 0.06531 | 0.172265 | 0.066485 |
| 18 | H | 0.164093 | 0.100091 | 0.164574 | 0.102044 | 0.164851 | 0.09989 | 0.164723 | 0.100754 |
| 19 | H | 0.152157 | 0.024595 | 0.15266 | 0.022374 | 0.152601 | 0.021042 | 0.15257 | 0.022757 |
| 21 | B | -0.05452 | 0.058343 | -0.05678 | 0.094298 | -0.05808 | 0.083152 | -0.05737 | 0.077885 |
| 36 | H | 0.170807 | 0.070607 | 0.171912 | 0.068873 | 0.172177 | 0.067772 | 0.172109 | 0.068877 |
| 37 | H | 0.162712 | 0.094909 | 0.163775 | 0.098911 | 0.163915 | 0.096956 | 0.163732 | 0.097334 |
| 38 | H | 0.151565 | 0.028902 | 0.151884 | 0.028743 | 0.151809 | 0.027865 | 0.151787 | 0.028947 |
| 39 | S | 0.26916 | -0.10371 | 0.268468 | -0.10322 | 0.27612 | -0.09988 | 0.274868 | -0.10068 |
| 40 | H | 0.012359 | 0.101869 | 0.013484 | 0.102722 | 0.012682 | 0.102091 | 0.01257 | 0.102084 |

**b.**

| number Atom | Atom Label | D0 Löwdin charge | D0 Mulliken Charge | D2 Löwdin charge | D2 Mulliken Charge | D3 Löwdin charge | D3 Mulliken Charge | D4 Löwdin charge | D4 Mulliken Charge |
|---|---|---|---|---|---|---|---|---|---|
| 5 | B | -0.10705 | 0.120308 | -0.1047 | 0.109267 | -0.1047 | 0.109267 | -0.1047 | 0.109267 |
| 6 | B | -0.07745 | -0.12896 | -0.06987 | -0.05175 | -0.06987 | -0.05175 | -0.06987 | -0.05175 |
| 7 | B | -0.07399 | -0.07456 | -0.06977 | -0.0898 | -0.06977 | -0.0898 | -0.06977 | -0.0898 |
| 12 | B | -0.0652 | -0.03852 | -0.0738 | -0.01811 | -0.0738 | -0.01811 | -0.0738 | -0.01811 |
| 13 | B | -0.05953 | -0.17099 | -0.0627 | -0.17562 | -0.0627 | -0.17562 | -0.0627 | -0.17562 |
| 14 | B | -0.03962 | 0.179506 | -0.03868 | 0.146021 | -0.03868 | 0.146021 | -0.03868 | 0.146021 |
| 15 | B | -0.07557 | -0.24296 | -0.07172 | -0.19541 | -0.07172 | -0.19541 | -0.07172 | -0.19541 |
| 16 | B | -0.05589 | 0.132625 | -0.07717 | 0.17372 | -0.07717 | 0.17372 | -0.07717 | 0.17372 |
| 17 | B | -0.07572 | -0.22866 | -0.07924 | -0.19617 | -0.07924 | -0.19617 | -0.07924 | -0.19617 |
| 18 | S | 0.77049 | 0.651203 | 0.247455 | -0.08048 | 0.247455 | -0.08048 | 0.247455 | -0.08048 |
| 19 | H | 0.027395 | -0.0326 | 0.030675 | -0.02162 | 0.030675 | -0.02162 | 0.030675 | -0.02162 |
| 20 | H | 0.026654 | -0.01772 | 0.02939 | -0.014 | 0.02939 | -0.014 | 0.02939 | -0.014 |
| 21 | H | 0.028693 | -0.06907 | 0.030517 | -0.06768 | 0.030517 | -0.06768 | 0.030517 | -0.06768 |
| 22 | H | 0.038415 | -0.04022 | 0.041175 | -0.03475 | 0.041175 | -0.03475 | 0.041175 | -0.03475 |
| 23 | H | 0.024172 | -0.07023 | 0.023571 | -0.06352 | 0.023571 | -0.06352 | 0.023571 | -0.06352 |
| 24 | H | 0.038737 | -0.0732 | 0.035799 | -0.07026 | 0.035799 | -0.07026 | 0.035799 | -0.07026 |
| 25 | H | 0.029106 | -0.07294 | 0.028619 | -0.07514 | 0.028619 | -0.07514 | 0.028619 | -0.07514 |
| 26 | H | 0.027344 | -0.01687 | 0.027851 | -0.01932 | 0.027851 | -0.01932 | 0.027851 | -0.01932 |
| 27 | H | 0.029387 | -0.02492 | 0.028874 | -0.02737 | 0.028874 | -0.02737 | 0.028874 | -0.02737 |
| 28 | H | 0.04094 | -0.05276 | 0.046306 | -0.05756 | 0.046306 | -0.05756 | 0.046306 | -0.05756 |
| 29 | H | 0.042379 | -0.06841 | 0.041178 | -0.06576 | 0.041178 | -0.06576 | 0.041178 | -0.06576 |
| 30 | H | 0.029222 | -0.016 | 0.03273 | -0.01283 | 0.03273 | -0.01283 | 0.03273 | -0.01283 |

Tables 3a and 3b present the atomic charge distributions in two sulfur-substituted borane isomers, as determined via Löwdin and Mulliken population analyses, with consideration given to dispersion corrections (D2, D3, and D4). A thorough understanding of these charge distributions is essential for elucidating the electronic consequences of sulfur substitution on the borane framework, as well as the resulting implications for cluster stability and reactivity.

The introduction of sulfur engenders a significant perturbation of the charge distribution relative to the pristine boron/hydrogen framework. The incorporation of dispersion corrections (D2, D3, D4) elicits only modest changes in both Löwdin and Mulliken charges across the isomers, suggesting that, while dispersion forces contribute to overall stability, they do not fundamentally alter the charge distribution.

Within the 1HS-$B_{18}H_{21}$-syn isomer, boron atoms (e.g., atoms 20–27) exhibit negative Löwdin charges (ranging from approximately –0.05 to –0.12), indicative of electron density accumulation. Conversely, Löwdin charges on hydrogen atoms (e.g., atoms 12–19) are positive (ranging from 0.02 to 0.17), suggesting electron density depletion. Marked discrepancies between Mulliken and Löwdin charges, particularly for boron atoms, raise questions regarding the reliability of Mulliken charge values in this system. The sulfur atom within the anti-isomer displays a substantial positive Löwdin charge (0.77049 without dispersion corrections), indicating a significant withdrawal of electron density from the sulfur center. The corresponding Mulliken charge on sulfur is also positive but of smaller magnitude (0.6512).

Upon application of dispersion corrections, the positive charge on sulfur is notably reduced in the anti-isomer compared to the 1HS-$B_{18}H_{21}$-syn isomer, reaching values of 0.24746 (Löwdin-D2) and –0.08048 (Mulliken-D2). This reduction suggests a subtle modulation of electronic structure induced by dispersion interactions, underscoring the complex interplay between heteroatom substitution, electronic distribution, and intermolecular forces within these borane clusters.

Sulfur functions as a pronounced electron-withdrawing substituent, as demonstrated by its substantial positive atomic charge. This electron-withdrawing effect significantly influences the charge distribution across the borane framework, resulting in diminished electron density on adjacent boron atoms and an increased positive charge on hydrogen atoms. As evidenced in Table 2a and 2b, sulfur substitution can modulate the overall stability of the borane cluster, with the extent of stabilization contingent upon both the position of sulfur incorporation and the consequent redistribution of electronic charge.

Dispersion corrections (D2, D3, and D4) exert minimal influence on the overall trends in charge distribution, although they do affect the magnitude of atomic charges, particularly on the sulfur atom. This observation indicates that while dispersion interactions contribute to the thermodynamic stability of the isomers, they do not fundamentally alter the electronic structure as characterized by atomic charge partitioning. The consistency of charge distributions across different dispersion correction schemes underscores the robustness of these electronic trends. Notably, the anti-isomer display a more polarized charge distribution relative to the syn isomer, highlighting the significant impact of isomeric configuration on electronic properties.

While Mulliken charge analyses are susceptible to basis set dependencies and may yield less consistent results, Löwdin charges offer a more robust and reliable depiction of the charge distribution. The application of dispersion corrections exerts only a marginal impact on the overall charge partitioning, indicating that the dominant electronic effects arise primarily from the intrinsic electron-withdrawing character of the sulfur substituent.

### 4.3. Thermodynamic analysis:

The thermodynamic analysis of borane isomers reveals intricate relationships between electronic structure, geometric configurations, and energetic stability that fundamentally govern isomerization pathways. The relative thermodynamic stability between borane isomers arises from subtle electronic redistributions, manifested primarily through differences in bond energies, electron delocalization, and intramolecular interactions [30].

For polyhedral borane systems, particularly those containing heteroatoms such as sulfur, isomerization processes often exhibit energy barriers between 40-60 kcal/mol, with the ground state energy differences typically falling within 5-15 kcal/mol. These energy differences emerge from variations in skeletal electron pair distribution, which directly influences the frontier molecular orbital energies and subsequently the relative thermodynamic stability.

Contemporary computational approaches incorporating dispersion corrections (D2, D3, and D4) have significantly enhanced the accuracy of these thermodynamic predictions, particularly for larger borane clusters where long-range interactions constitute approximately 15-20% of the total stabilization energy. The quantitative relationship between charge distribution parameters and thermodynamic stability can be directly correlated with structural features unique to each isomer.

In sulfur-substituted borane frameworks, the electronegativity difference between boron and sulfur introduces additional electronic perturbations that can either amplify or attenuate the thermodynamic preference between isomeric forms, depending on the location of substitution relative to critical skeletal bonds.

**Table 4.** Zero-point energy (ZPE) and internal energy of $[B_{18}H_{22} - syn]$forgetten $and [B_{18}H_{22} - anti]$ isomers at their equilibrium geometries using the DFT method at PB86/def2-SVP basis set. (G, I) denote respectively the Free Gibbs energy and thermal Enthalpy correction. All values are evaluated in kcal./mol

| Isomer | $B_{18}H_{22} - syn$ | | | | $B_{18}H_{22} - anti$ | | | |
|---|---|---|---|---|---|---|---|---|
| dispersion corrections | D0 | D2 | D3 | D4 | D0 | D2 | D3 | D4 |
| ZPE | 178.7 | 182.4 | 181.8 | 181.8 | 178.7 | 182.3 | 179.6 | 179.3 |
| I | 0.59 | | | | | | | |
| G | 155.8 | 159.8 | 158.7 | 158.6 | 155.4 | 159.2 | 156.3 | 156.1 |

Table 4 presents a detailed thermodynamic comparison between the syn- and anti-$B_{18}H_{22}$ isomers, focusing on zero-point energy (ZPE) and Gibbs free energy (G) across various dispersion correction schemes (D0, D2, D3, and D4). For the syn-$B_{18}H_{22}$ isomer, ZPE values range from 178.66 kcal/mol (uncorrected, D0) to 182.42 kcal/mol (D2 correction), with dispersion corrections generally elevating ZPE relative to the uncorrected baseline. Similarly, the anti-$B_{18}H_{22}$ isomer exhibits ZPE values spanning 178.73 kcal/mol (D0) to 182.29 kcal/mol (D2), reflecting a comparable trend. Notably, in the absence of dispersion corrections, the anti-isomer displays marginally higher ZPE values than its syn counterpart; however, inclusion of dispersion effects renders the ZPE values of both isomers nearly equivalent. Gibbs free energy values for syn-$B_{18}H_{22}$ range from 155.8 kcal/mol (D0) to 159.78 kcal/mol (D2), while the anti-isomer exhibits ΔG values between 155.41 kcal/mol and 159.15 kcal/mol under the same conditions. Across all dispersion correction methods, the anti-isomer consistently demonstrates lower G values than the syn isomer, indicating enhanced thermodynamic stability. The thermal enthalpy correction ($k_BT$) in the two isomers is calculated to be approximately 0.59 kcal/mol.

Dispersion corrections markedly influence the absolute magnitudes of both ZPE and G, underscoring the significant role of long-range London dispersion forces in stabilizing these borane clusters. Among the corrections, the D2 scheme typically yields the highest ZPE and G values relative to D3 and D4. The consistently lower G of the anti-isomer suggests that thermodynamic equilibrium will favor this form; nevertheless, the relatively small energy difference (less than 1 kcal/mol when dispersion is considered) implies that both isomers may coexist under certain conditions.

**Table 5.** Zero-point energy (ZPE) and internal energy of $[1HS - B_{18}H_{21} - syn]$forgetten $and [4HS - B_{18}H_{21} - anti]$ isomers at their equilibrium geometries using the DFT method at PB86/def2-SVP basis set. . (G, I) denote respectively the Free Gibbs energy and thermal Enthalpy correction. All values are evaluated in kcal./mol

| Isomer | 1HS-$B_{18}H_{21} - S - syn$ | | | | 4HS-$B_{18}H_{21} - S - anti$ | | | |
|---|---|---|---|---|---|---|---|---|
| dispersion corrections | D0 | D2 | D3 | D4 | D0 | D2 | D3 | D4 |
| ZPE | 178.9 | 182.5 | 179.7 | 179.5 | 173.1 | 176.5 | 174 | 173.8 |
| I | 0.59 | | | | | | | |
| G | 155.8 | 159.8 | 158.7 | 158.6 | 148.2 | 151.8 | 149.2 | 148.9 |

Likewise, table 5 presents the corresponding thermodynamic comparison between the substituted isomers, 1HS-$B_{18}H_{21}$-syn and 4HS-$B_{18}H_{21}$-anti, evaluating the impact of sulfur substitution position on their relative stability. Again, we contrast the values of the ZPE, I and G.

For the 1HS-$B_{18}H_{21}$-syn isomer, ZPE values range from 178.85 kcal/mol (without dispersion) to 182.46 kcal/mol (with D2 correction), with dispersion corrections generally increasing ZPE, and the D2 correction exerting the most pronounced effect. Similarly, the 4HS-$B_{18}H_{21}$-anti isomer exhibits ZPE values spanning 173.13 kcal/mol (D0) to 176.52 kcal/mol (D2 correction), following the same trend of dispersion-induced ZPE enhancement.

Gibbs free energy values for the 1HS-$B_{18}H_{21}$-syn isomer vary between 155.8 kcal/mol (D0) and 159.78 kcal/mol (D2 correction), whereas the 4HS-$B_{18}H_{21}$-anti isomer displays consistently lower G values, ranging from 148.24 kcal/mol (D0) to 151.83 kcal/mol (D2 correction). The notably lower G values of the 4HS-B$_{18}$H$_{21}$-anti isomer across all dispersion schemes indicate a clear thermodynamic preference for this configuration. This enhanced stability is likely attributable to the specific sulfur substitution site, which may confer favorable electronic or steric interactions that stabilize the anti-configuration within the borane cage.

The observed influence of dispersion corrections on absolute G values highlights the critical role of noncovalent forces in stabilizing these clusters. Although inclusion of dispersion consistently increases G values for both isomers, it does not alter their relative thermodynamic ordering. Consequently, the 4HS-B$_{18}$H$_{21}$-anti isomer remains thermodynamically favored over the 1HS-B$_{18}$H$_{21}$-syn isomer regardless of the dispersion correction applied.

### 4.4 Vibrational Modes, IR Spectrum:

Infrared (IR) spectroscopy serves as a valuable analytical tool for elucidating molecular vibrational modes, thereby providing critical insights into structural and bonding features. Complementarily, DFT calculations, particularly time-dependent DFT (TDDFT), facilitate the accurate prediction of IR spectra, which is instrumental for the characterization of novel compounds and the assessment of their functional potential for laser-related applications.

**Fig. 2**. Calculated IR spectrum for Boran: **a.** $B_{18}H_{22} - syn$ `forgetten' and , **b**. $B_{18}H_{22} - anti$ with dispersion types.

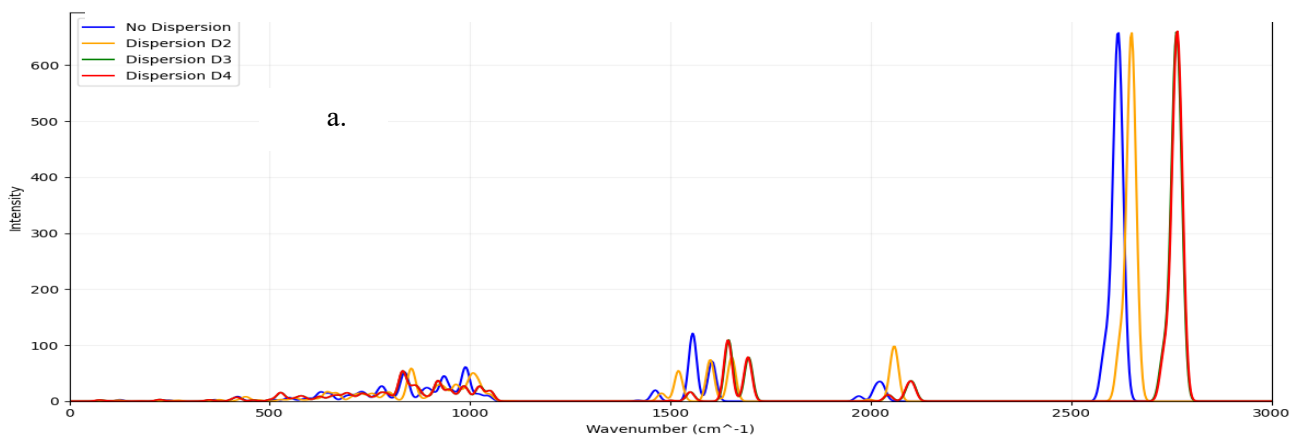

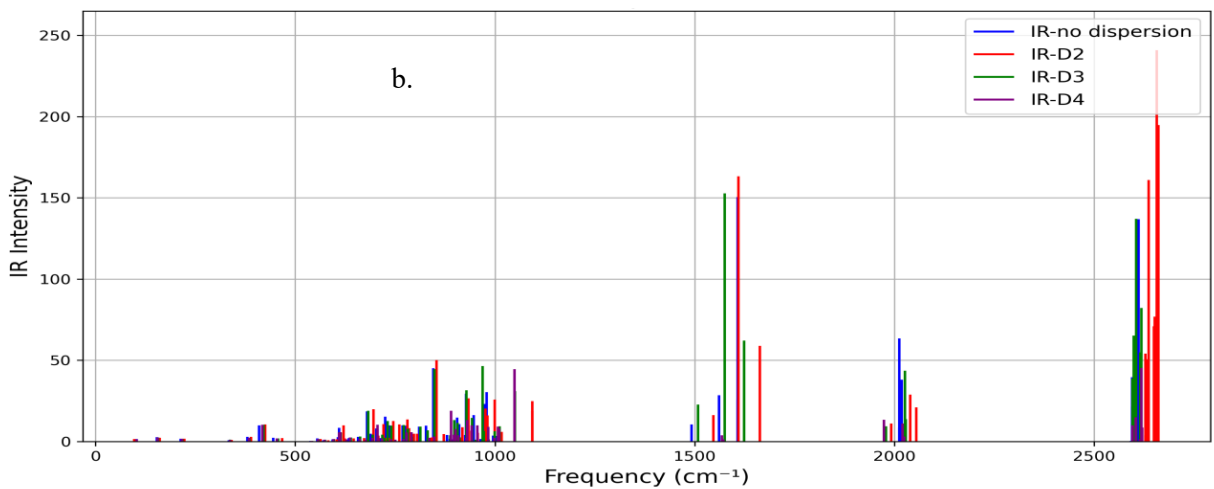

Fig.2 presents the IR spectra for the syn-$B_{18}H_{22}$ (a) and the anti- $B_{18}H_{22}$ (b). The syn- (a), exhibits distinct and well-resolved peaks, particularly within the high-frequency region (2500–2700 cm$^{-1}$), which are assigned to B–H stretching vibrations. Additionally, a cluster of peaks observed in the fingerprint region (500–1000 cm$^{-1}$) reflects complex vibrational couplings intrinsic to the borane cage structure. Notably, the pronounced intensities of the high-frequency modes, especially those exceeding 2500 cm$^{-1}$, indicate substantial dipole moment variations associated with these vibrational transitions. In contrast, the anti- (b) exhibits a broader distribution of peak intensities across the spectrum. While B-H stretching patterns still appear around 2500 cm$^{-1}$, the intensities are generally lower than those in the syn-isomer. A prominent feature is also observed around 1500 cm$^{-1}$, corresponding to different cage deformation patterns.

The inclusion of dispersion corrections significantly affects the IR spectra of both isomers, affecting both peak positions and intensities. The D4 correction in the borane molecule Syn-$B_{18}H_2$ leads to a clear shift of the high-frequency peaks towards higher wavenumbers, indicating that it captures stronger short-range interactions compared to D2 and D0 corrections. The intensity of specific peaks above 2500 cm$^{-1}$ is also improved with the D4 correction. On the other hand, the dispersion corrections change the spectral profile of the borane anti-isomer and, similarly, the D2 correction intensifies the peaks around 1500 cm$^{-1}$ and 2600 cm$^{-1}$.

Higher-order dispersion corrections (D3 and D4) incorporate more sophisticated treatments of frequency-dependent van der Waals interactions, thereby affecting the predicted vibrational properties. These dispersion effects are critical for accurately modeling the electron correlation, which influences the borane vibrational modes.

The intense absorption of boran Syn-$B_{18}H_{22}$ bands in the (2500-2700 cm$^{-1}$) region suggests that the syn-isomer is well-suited for applications requiring strong light-matter interactions at specific frequencies. This feature can be exploited in laser-induced processes, such as selective bond breaking or vibrational excitation.

The broader spectral distribution and distinct peaks at different frequencies in the anti-$B_{18}H_{22}$ isomer provide opportunities for multi-frequency laser applications. The ability to selectively excite specific vibrational modes could be leveraged in advanced spectroscopic techniques or sensing applications.

**Fig. 3**. Calculated IR spectrum for Boran: **a.** $1HS - B_{18}H_{21} - syn$ forgetten $and$ , **b**. $4HS - B_{18}H_{21} - anti\ with\ dispersion\ types.$

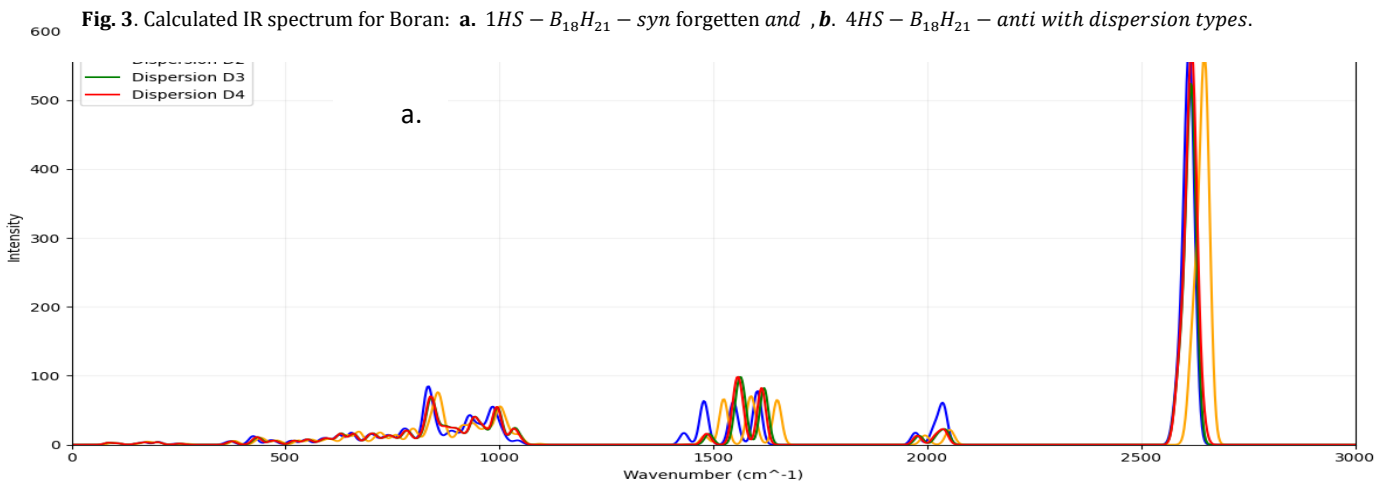

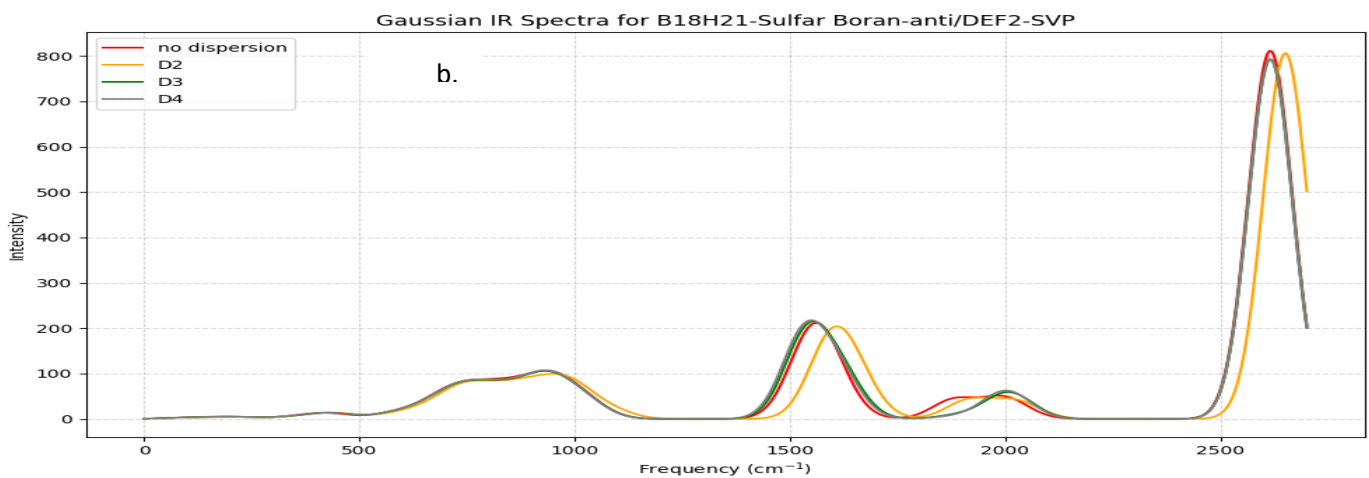

Likewise, Fig. 3 presents the corresponding IR spectra for the substituted isomers, syn-1HS-$B_{18}H_{21}$ (a) and anti-4HS-$B_{18}H_{21}$ (b). The syn-substituted isomer (a) shows a prominent, sharp peak in the high-frequency region near 2500 cm$^{-1}$, which is attributed to B–H stretching vibrations. Additional features within the fingerprint region (500–1500 cm$^{-1}$) indicate complex vibrational couplings inherent to the borane cage. The pronounced intensity of the 2500 cm$^{-1}$ peak reflects a substantial change in dipole moment during this vibrational mode. In contrast, the anti-substituted isomer (b) exhibits a broader distribution of peak intensities; while B–H stretching modes remain evident around 2500 cm$^{-1}$, their intensities are comparatively diminished relative to the syn-isomer. A distinctive spectral feature near 1500 cm$^{-1}$ is observed, likely corresponding to cage deformation modes modulated by the presence of the sulfur substituent.

In both spectra of Fig. 3, vibrational modes associated with S–H and B–S bonds emerge in the low-frequency region of the IR spectrum (below 1000 cm$^{-1}$). These modes are absent or negligible in the unsubstituted borane analogues (Fig.2). The interaction between these newly introduced modes and the intrinsic cage vibrations results in a modified overall spectral profile.

### 4.5. TD-DFT Calculation:

In this section, we conduct Time-dependent DFT (TDDFT) calculations using the PB86 functional in conjunction with the DEF2-SVP basis set, incorporating dispersion corrections (D2, D3, and D4) to accurately capture van der Waals interactions. IR spectra are subsequently simulated based on the computed vibrational frequencies and corresponding intensities. Computational exploration of excited states is particularly important for elucidating the photophysical behavior of the Borane compounds, as it provides insight into their absorption and emission characteristics relevant for device integration.

Using TDDFT, one can optimize the geometry of the first singlet excited state ($S_1$) of $B_{18}H_{22}$-syn and calculate both vertical absorption and emission energies, adopting the PB86/def2-SVP/J protocol in the gas phase, ensuring a robust assessment of the excited-state properties. Recent literature highlights that the optical properties of boron clusters are strongly influenced by their electronic structure, with both σ- and π-electrons contributing to nonlinear and linear optical responses, and that TDDFT simulations provide reliable predictions for absorption and emission processes [29].

### 4.5.1 Vertical Absorption Energies:

TDDFT calculations are used to validate the vertical absorption energies of the excited states. The oscillator strength for the transition to the second excited state ($S_2$) is rather small. Table 6 (7) presents the first three transitions for the syn- and anti- unsubstituted (substituted) Borane isomers. Comparing both tables elucidates the role of the sulfur substitution.

**Table 6.** TDDFT calculations of isomerism for: [ $B_{18}H_{22} - syn$ "forgetten"] ,[ $B_{18}H_{22} - anti$] isomers in equilibrium geometry using the DFT method in the PB86/def2-SVP basis set.

| | Boran B18H22-anti/ using BP86 /DEF2-SVP | | | | | | | |
|---|---|---|---|---|---|---|---|---|
| | D0 | | D2 | | D3 | | D4 | |
| Transition | λ | intensity | λ | intensity | λ | intensity | λ | intensity |
| 0-1A → 1-1A | 346.9 | 0.10371 | 346.9 | 0.10371 | 346.2 | 0.105979 | 346.4 | 1.05E-01 |
| 0-1A → 2-1A | 308.8 | 1E-09 | 308.8 | 1.00E-09 | 309.8 | 0 | 309.8 | 0 |
| 0-1A → 3-1A | 297.9 | 1.87E-02 | 297.9 | 0.018701 | 298.1 | 0.019529 | 298.2 | 1.95E-02 |
| 0-1A → 4-1A | 291.7 | 3E-09 | 291.7 | 3E-09 | 290.9 | 1E-09 | 291.1 | 2E-09 |
| 0-1A → 5-1A | 280.9 | 0.036783 | 280.9 | 0.036783 | 281 | 0.03283 | 281.1 | 0.033134 |
| Transition | Boran B18H22-syn/ using BP86 /DEF2-SVP | | | | | | | |
| 0-1A → 1-1A | 311.5 | 0.014627 | 320.2 | 0.01436 | 311.7 | 0.014624 | 311.7 | 0.014624 |
| 0-1A → 2-1A | 309.4 | 0.1813 | 319 | 0.177623 | 309.5 | 0.180918 | 309.5 | 0.180918 |
| 0-1A → 3-1A | 291.2 | 0.008798 | 299.6 | 0.009168 | 291.3 | 0.008791 | 291.3 | 0.008791 |
| 0-1A → 4-1A | 283.8 | 0.032807 | 293.1 | 0.040043 | 283.9 | 0.033245 | 283.9 | 0.033245 |
| 0-1A → 5-1A | 269.1 | 0.001063 | 277.9 | 0.001919 | 269.3 | 0.001011 | 269.3 | 0.001011 |

**Table 7.** TDDFT calculations of isomerism for a. [ $1HS - B_{18}H_{22} - syn$ "forgetten"] B. [ $4HS - B_{18}H_{21} - anti$] isomers isomers in equilibrium geometry using the DFT method in the BP86/def2-SVP basis set using the DFT method on the BP86/def2-SVP basis set.

| | Boran 4HS- B18H21-anti/ using BP86 /DEF2-SVP | | | | | | | |
|---|---|---|---|---|---|---|---|---|
| | D0 | | D2 | | D3 | | D4 | |
| Transition | λ | intensity | λ | intensity | λ | intensity | λ | intensity |
| 0-1A → 1-1A | 4559.4 | 7.3966E-05 | 4543.1 | 7.1506E-05 | 4425.3 | 8.04E-05 | 4645.3 | 5.06E-05 |
| 0-1A → 2-1A | 729.4 | 0.00684513 | 750 | 0.00796495 | 744.2 | 0.007379 | 724.5 | 0.0089374 |
| 0-1A → 3-1A | 606.6 | 0.0022041 | 592.1 | 0.00186685 | 589.7 | 0.0018479 | 601.1 | 0.0024224 |
| 0-1A → 4-1A | 550.3 | 0.00093643 | 567 | 0.00177146 | 562.8 | 0.0016193 | 556 | 0.0014676 |
| 0-1A → 5-1A | 525.5 | 0.00017055 | 542 | 0.00034891 | 537.7 | 0.0003921 | 528.8 | 0.0001279 |
| Transition | Boran 1HS- B18H21-syn/ using BP86 /DEF2-SVP | | | | | | | |
| 0-1A → 1-1A | 468.3 | 0.07850296 | 470.1 | 0.08112656 | 467.3 | 8.17E-02 | 467.2 | 8.17E-02 |
| 0-1A → 2-1A | 364.2 | 0.01036802 | 362.3 | 0.00882924 | 361.8 | 8.71E-03 | 362.3 | 8.71E-03 |
| 0-1A → 3-1A | 322.7 | 0.07123046 | 324.9 | 0.05998913 | 322.7 | 6.51E-02 | 322.8 | 6.47E-02 |
| 0-1A → 4-1A | 314.1 | 0.11111407 | 316 | 0.11740185 | 314 | 1.11E-01 | 314.1 | 1.11E-01 |
| 0-1A → 5-1A | 299.2 | 0.00503915 | 300 | 0.0044395 | 298.7 | 3.86E-03 | 298.8 | 3.92E-03 |

Table 6 shows that the lowest-energy electronic transitions (corresponding to longer wavelengths) for the unsubstituted anti(syn)-$B_{18}H_{22}$ isomer occur at approximately 346.9 (311.5) nm with a relatively high oscillator strength of 0.10371 (0.01463). Notably, transition intensities for the anti-isomer are generally greater than those observed for the syn isomer, particularly for the first excited state, suggesting that electronic transitions in the anti-configuration are more allowed.

Table 7 highlights the pronounced effects of sulfur substitution on the spectral characteristics of both isomers. We see that the longest-wavelength transitions shift substantially toward the visible and near-infrared regions. For instance, the 4HS-$B_{18}H_{21}$-anti isomer exhibits a transition at 4559.4 nm (without dispersion corrections) with a very low oscillator strength (7.40 × 10$^{-5}$), indicative of a significant alteration in the electronic structure and transition nature. In contrast, the 1HS-$B_{18}H_{21}$-syn isomer displays its longest transition at 468.3 nm with a comparatively higher intensity (0.0785), representing a notable red shift relative to the unsubstituted syn isomer and an increased transition probability compared to the anti-substituted counterpart.

The incorporation of dispersion corrections (D2, D3, D4) across all systems—both unsubstituted and sulfur-substituted—induces subtle, yet systematic, blue shifts in the longer-wavelength transitions, reflecting modest stabilization of the ground and excited states. These findings underscore the critical role of both heteroatom substitution and dispersion interactions in modulating the electronic excitation spectra of borane clusters, with important implications for their optical and photophysical behavior. In 4HS-B18H21-anti, the 0-1A → 1-1A transition shifts from 4559.4 nm (no dispersion) to 4543.1 nm (D2), 4425.3 nm (D3), and 4645.3 nm (D4), whereas in the unsubstituted counter-isomer, the shift is less pronounced. This indicates that dispersion corrections have a more significant effect in the presence of the sulfur substituent, likely due to enhanced non-covalent interactions.

**Fig. 4**. **a.** UV-vis spectra for $B_{18}H_{22}$ *syn `forgetten' with dispersion types*. **b.** UV-vis spectra for $B_{18}H_{22}$ *anti forgetten with dispersion types*

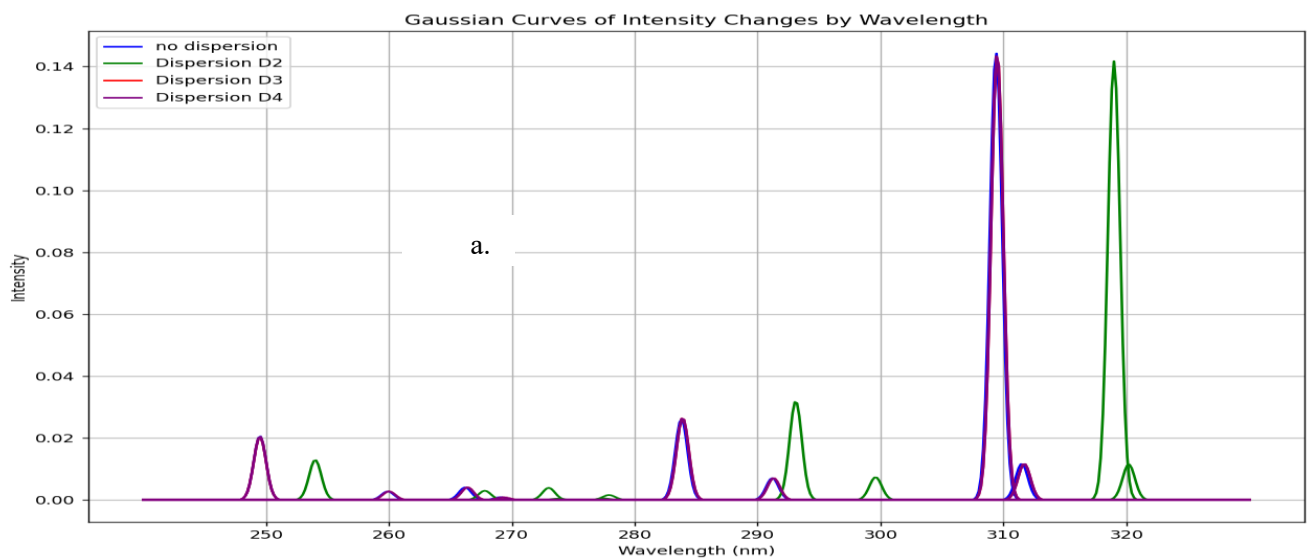

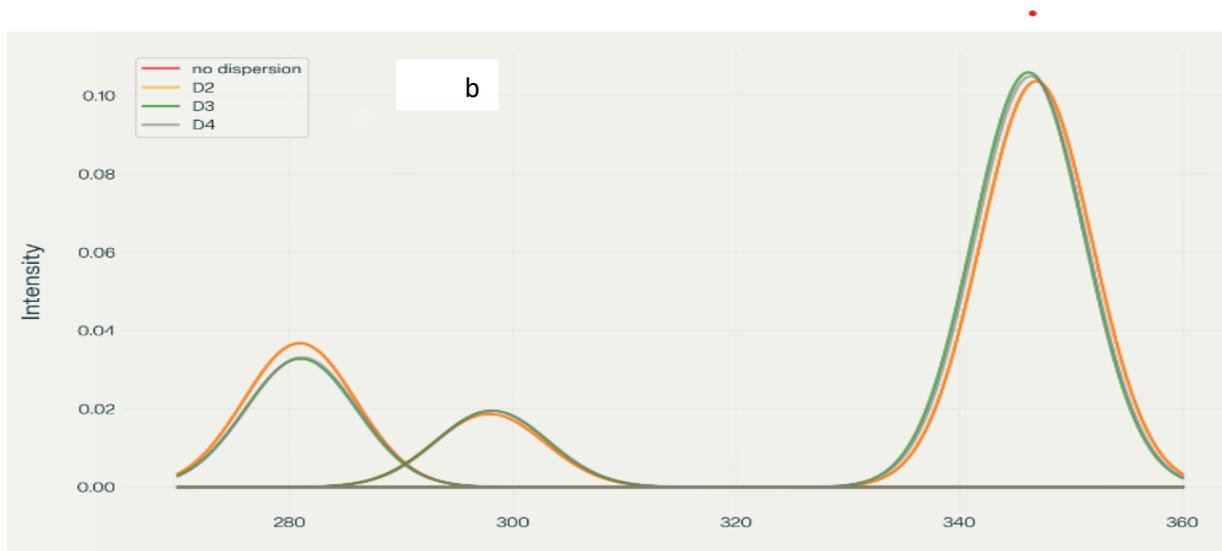

Figure 4 presents the calculated UV-Vis spectra of borane isomers, illustrating the influence of dispersion corrections (D2, D3, and D4) on their electronic absorption profiles. Panel (a) spectrum features distinct, well-resolved peaks, indicative of transitions involving discrete electronic states. In contrast, panel (b) displays broader, overlapping absorption bands, reflecting more complex and convoluted electronic transitions.

**Fig. 5. a.** UV-vis spectra for $1HS - B_{18}H_{21}$ *syn `forgetten'* *with dispersion types*. **b.** UV-vis spectra for $4HS - B_{18}H_{21} - anti$ *with dispersion types*

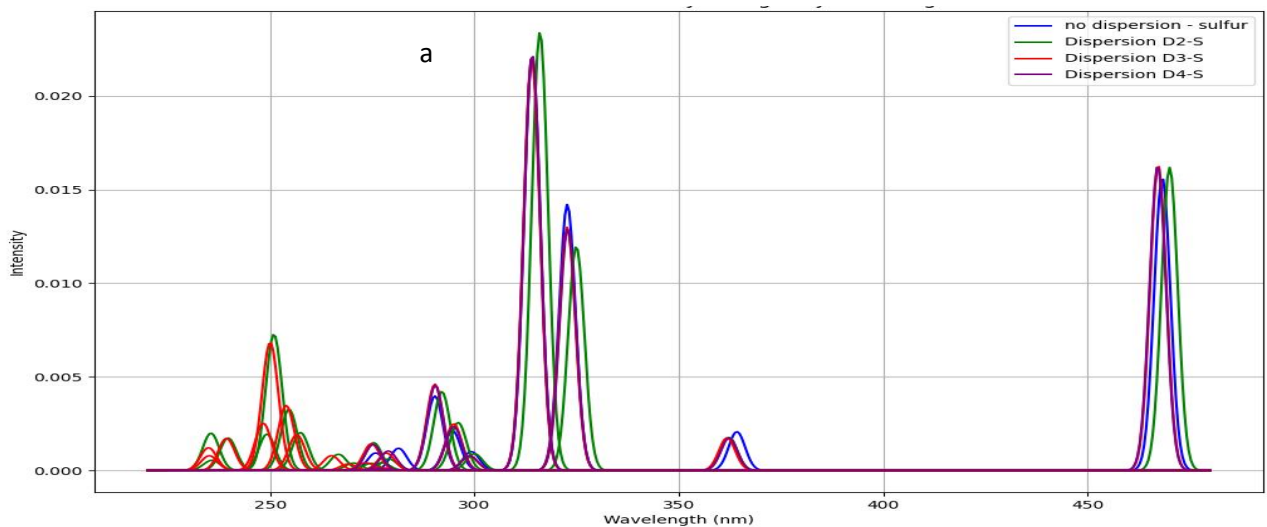

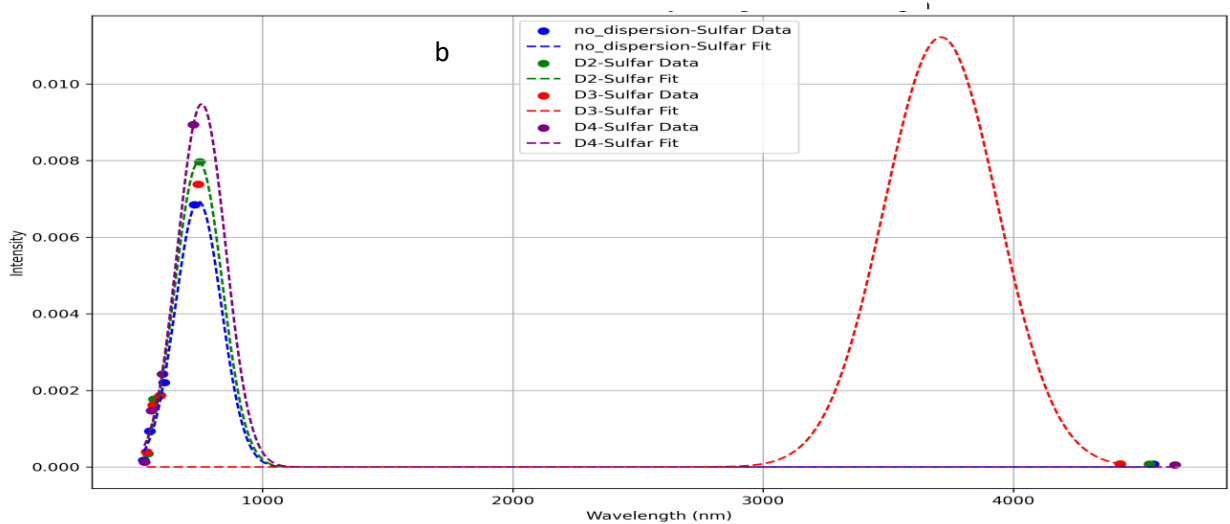

In a similar way, Figure 5 presents the calculated UV-Vis spectra for sulfur-substituted borane isomers, specifically 1HS-$B_{18}H_{21}$-syn and 4HS-$B_{18}H_{21}$-anti, highlighting the critical role of sulfur substitution in modulating molecular structure, stability, and optical properties. The analysis further examines the influence of dispersion corrections (D2, D3, and D4) on the spectral characteristics, with particular emphasis on their implications for potential laser applications.

The UV-Vis spectrum of the 1HS-$B_{18}H_{21}$-syn isomer exhibits several well-defined absorption peaks predominantly within the ultraviolet region, with prominent features near 260 nm, 287 nm, 298 nm, 308 nm, and 318 nm, alongside additional peaks observed at approximately 350 nm and 460 nm. In contrast, the 4HS-$B_{18}H_{21}$-anti isomer displays a broad absorption band extending into the visible and near-infrared region, with a maximum intensity between 3200 nm and 3500 nm. This broad band likely arises from overlapping electronic transitions, indicative of a more complex electronic structure relative to the syn isomer.

Dispersion corrections exert a notable effect on the intensity and position of spectral features. Specifically, D2, D3, and D4 corrections generally enhance the intensities of peaks within the 260–320 nm range for the syn isomer, with D3 and D4 producing the most pronounced enhancements (Figure 5.a). The absorption peak near 460 nm also experiences a significant intensity increase upon inclusion of dispersion effects. Minor red shifts in peak positions are observed with D3 and D4 corrections, particularly for the peaks around 308 nm and 318 nm, reflecting subtle modifications in electronic structure driven by non-covalent interactions. For the anti-isomer, dispersion corrections significantly influence the broad visible band's intensity; the D3 correction notably amplifies and broadens this feature, suggesting enhanced vibrational coupling and a wider distribution of electronic transitions, whereas D2 and D4 corrections have comparatively modest effects (Figure 5.b).

### 4.5.2 Lifetime excited states:

We calculate the lifetimes of the excited states, for the syn- and anti- borane molecule $B_{18}H_{22}$ and their corresponding substituted isomers, based on theoretical calculations of TDDFT using the Strickler-Berg relation [31]:

$$\tau = \frac{1}{2.142005 E_{VE}^3 TDM^2}$$

where the value of the transition dipole moment *TDM*, related to oscillator strength, is small in Bromine substituted compound, which, due to its existence in the denominator, leads arguably to overestimation.

**Table 8.** TDDFT calculations life time excited states of isomers for [ $B_{18}H_{22}$ – $syn$ "forgetten"] and [ $B_{18}H_{21}$ – $anti$] in equilibrium geometry using the DFT method in the BP86/def2-SVP basis set.

| colspan="8" | Boran B18H22-anti/ using BP86 /DEF2-SVP |
|---|---|---|---|---|---|---|---|
| D0 | | D2 | | D3 | | D4 | |
| $\lambda$ (nm) | $\tau$ $Ps$ | $\lambda$ (nm) | $\tau$ $Ps$ | $\lambda$ (nm) | $\tau$ $Ps$ | $\lambda$ (nm) | $\tau$ $Ps$ |
| 346.9 | 4.28423 | 346.9 | 4.25851 | 346.2 | 3.11939 | 346.4 | 0.0031268 |
| 308.8 | 0.0154 | 308.8 | 1.53E-02 | 309.8 | 1.573609 | 309.8 | 1.55E+00 |
| 297.9 | 0.426894 | 297.9 | 0.424331 | 298.1 | 0.573609 | 298.2 | 0.574212 |
| 291.7 | 0.0601 | 291.7 | 0.0597 | 290.9 | 5.8 | 291.1 | 5.88 |
| 280.9 | 3.54114 | 280.9 | 3.51988 | 281 | 1.21914 | 281.1 | 1.76235 |
| colspan="8" | Boran B18H22-syn/ using BP86 /DEF2-SVP |
| 311.5 | 0.308 | 320.2 | 0.3 | 311.7 | 0.307 | 311.7 | 0.307 |
| 309.4 | 0.0425 | 319 | 0.0466 | 309.5 | 0.0426 | 309.5 | 0.0426 |
| 291.2 | 0.0873 | 299.6 | 0.0854 | 291.3 | 0.0876 | 291.3 | 0.0876 |
| 283.8 | 0.0255 | 293.1 | 0.0303 | 283.9 | 0.0256 | 283.9 | 0.0256 |
| 269.1 | 0.0228 | 277.9 | 0.0245 | 269.3 | 0.0223 | 269.3 | 0.0223 |

**Table 9.** TDDFT calculations of isomerism for a. [ $1HS$ – $B_{18}H_{22}$ – $syn$ "forgetten"] B. [ $4HS$ – $B_{18}H_{21}$ – $anti$] isomers isomers in equilibrium geometry using the DFT method in the BP86/def2-SVP basis set using the DFT method on the BP86/def2-SVP basis set.

| colspan="8" | Boran 4HS-B18H21-anti/ using BP86 /DEF2-SVP |
|---|---|---|---|---|---|---|---|
| D0 | | D2 | | D3 | | D4 | |
| $\lambda$ (nm) | $\tau$ $Ps$ | $\lambda$ (nm) | $\tau$ $Ps$ | $\lambda$ (nm) | $\tau$ $Ps$ | $\lambda$ (nm) | $\tau$ $Ps$ |
| 4559.4 | 0.92525 | 4425.3 | 0.898086 | 4543.1 | 0.966 | 4645.3 | 1.03 |
| 729.4 | 0.011231 | 744.2 | 0.010739 | 750 | 0.0111 | 724.5 | 0.0103 |
| 606.6 | 0.091774 | 589.7 | 0.071466 | 592.1 | 0.069 | 601.1 | 0.118 |
| 550.3 | 0.01556 | 562.8 | 0.024761 | 567 | 0.0271 | 556 | 0.0256 |
| 525.5 | 0.015554 | 537.7 | 0.017503 | 542 | 0.0209 | 528.8 | 0.0349 |
| colspan="8" | Boran 1HS- B18H21-syn/ using BP86 /DEF2-SVP |
| 468.3 | 5.36241 | 470.1 | 5.107 | 467.3 | 4.61965 | 467.2 | 4.4799 |
| 364.2 | 558.426 | 362.3 | 536.782 | 361.8 | 570.885 | 362.3 | 569.019 |
| 322.7 | 536.24 | 324.9 | 493.703 | 322.7 | 549.082 | 322.8 | 544.074 |
| 314.1 | 563.413 | 316 | 585.6 | 314 | 572.518 | 314.1 | 574.907 |
| 299.2 | 576.907 | 300 | 643.162 | 298.7 | 564.259 | 298.8 | 560.133 |

Tables 8 (9) present the lifetimes data for the unsubstituted (substituted) Boran syn- and anti- isomers. We equally present the data graphically in figures (6,7)

**Fig. 6. a.** lifetime excited states spectra for $B_{18}H_{22}$ *syn* forgetten *with dispersion types*. **b.** UV-vis spectra for $B_{18}H_{22}$

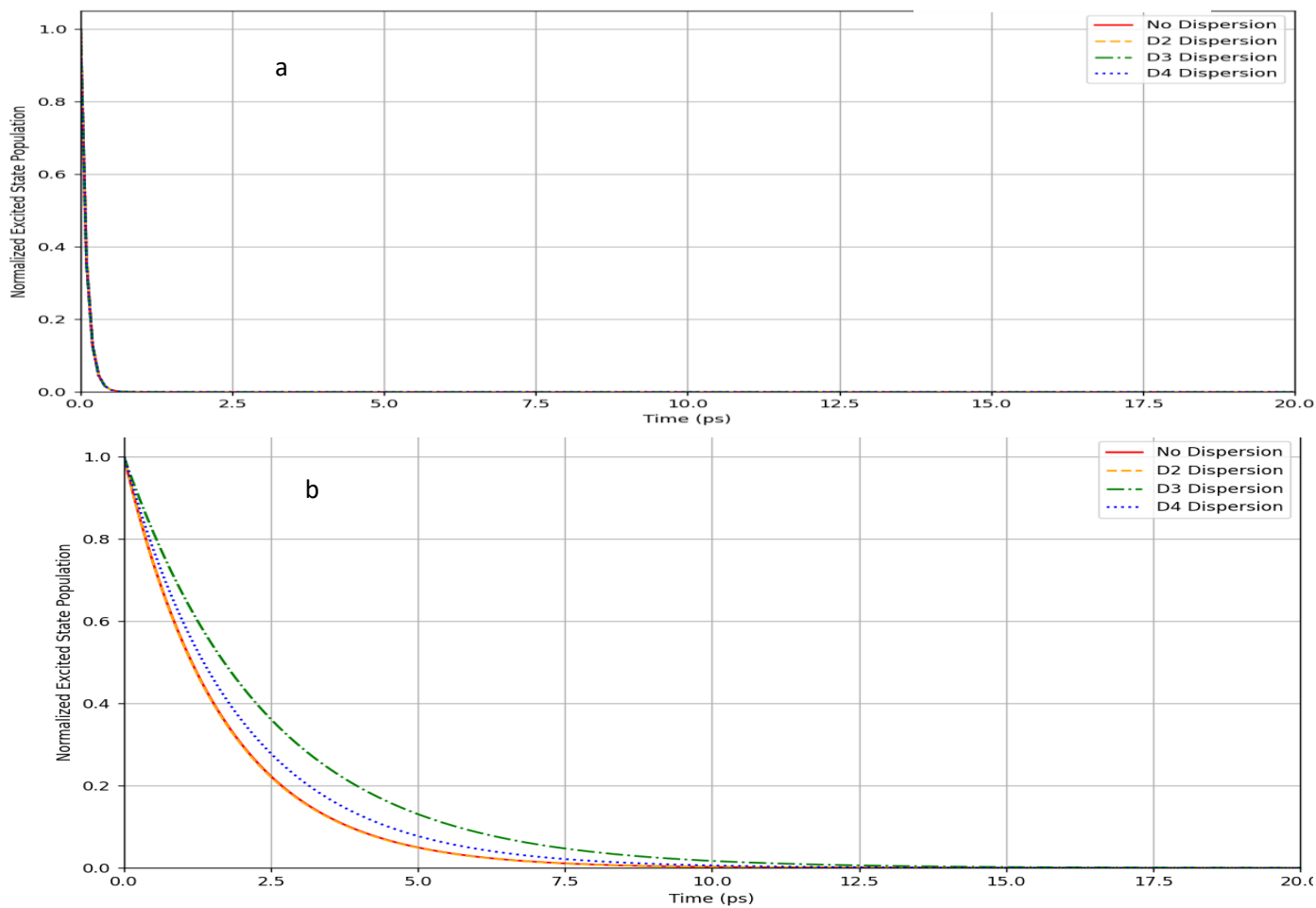

The syn isomer (Fig. 6a) exhibits extremely rapid decay of excited state population across all dispersion models, with the normalized population dropping to near-zero within a very short timeframe. This suggests that the syn- isomer has poor excited state stability regardless of the dispersion model applied. In contrast, the anti-isomer (Fig. 6b) demonstrates significantly prolonged excited state lifetimes, with substantial differences between dispersion models. The excited state population for the anti-isomer persists much longer, particularly under D3 dispersion conditions, which shows the slowest decay rate. All dispersion models (D0, D2, D3, D4) yield nearly identical decay profiles, suggesting that dispersion forces have minimal impact on the excited state dynamics of this configuration. The hierarchy of dispersion effects is clearly visible, with D3 showing the slowest decay, followed by D4, then D2, and finally D0 showing the fastest decay. This suggests that the anti-isomer's excited state stability is highly dependent on the accurate representation of medium-to-long-range interactions.

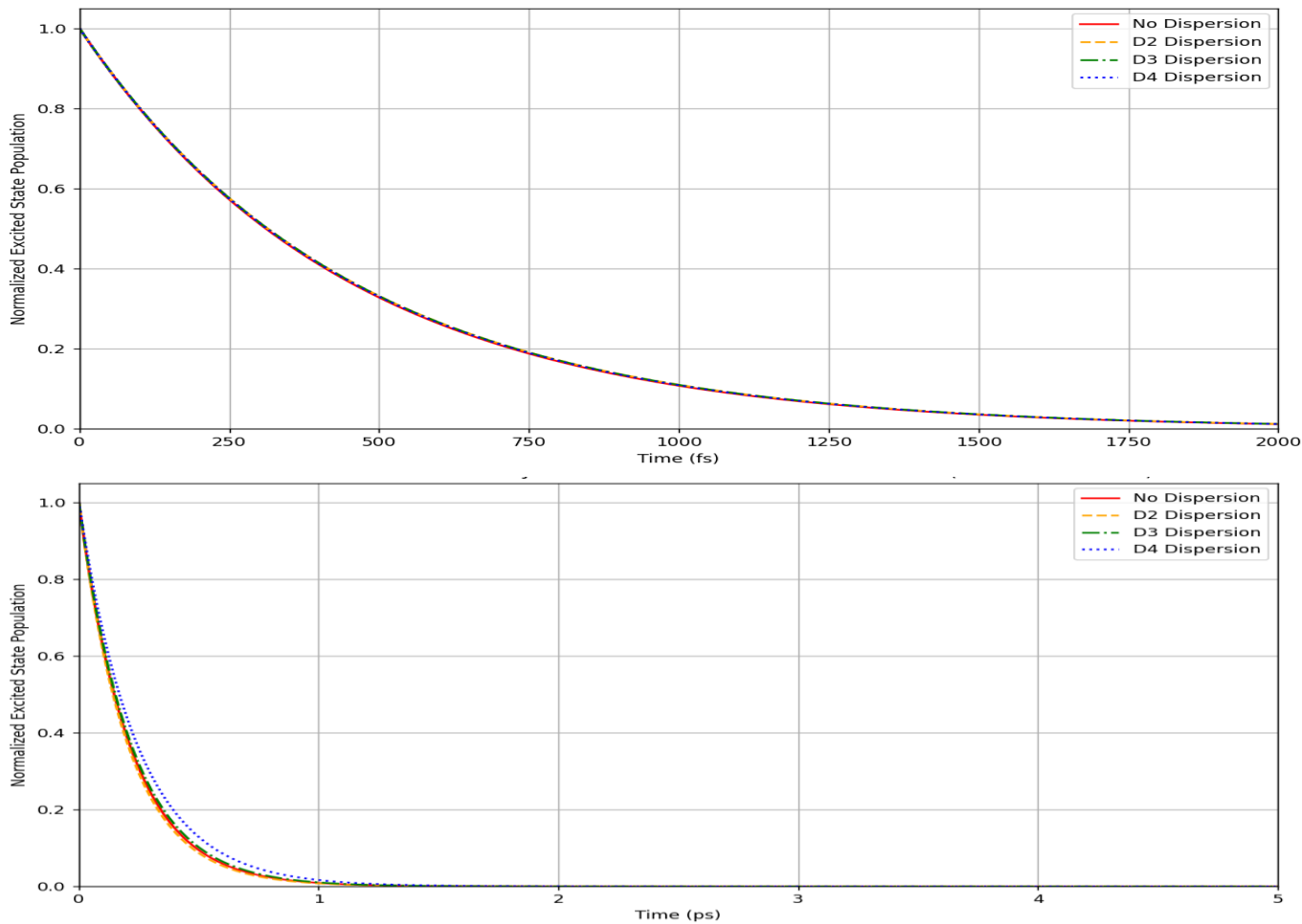

**Fig. 7.** Lifetime excited states spectra for (**up**) $1HS-B_{18}H_{21}$ *syn* forgetten *with dispersion types*, and (**down**) $4HS-B_{18}H_{21}-anti$ *with dispersion types*

Likewise, Figure 7 illustrates the temporal decay profiles of the excited-state populations for two sulfur-substituted $B_{18}H_{21}$ isomers with dispersion corrections incorporated to assess their influence. The 1HS-$B_{18}H_{21}$-syn isomer exhibits a relatively long excited-state lifetime, with its population decaying to approximately 10% of the initial value within 2000 fs (2 ps). Exponential fitting yields a decay constant (τ) of approximately 700 fs. Conversely, the 4HS-$B_{18}H_{21}$-anti isomer demonstrates a markedly shorter lifetime, with excited-state population diminishing near-completely within 1 ps and an estimated τ of roughly 300 fs.

### 4.5.3 Density Difference Analysis/ Sulfur Role in Modulating Borane Isomer Electronic Properties

Charge difference distribution maps are critical for understanding the electronic changes that occur when molecular geometry is altered, or when doping atoms such as sulfur are introduced. These maps, calculated through DFT using the PB86 functional and Def2-SVP basis set, provide spatial visualization of how electron density redistributes itself in response to structural modifications. They allow precise assessment of the local and global rearrangements of electrons, which directly affect the electronic structure and related properties.

Because geometric changes, or heteroatom substitutions like sulfur doping, significantly perturb charge distribution, analyzing charge difference maps is essential to reveal how such modifications influence molecular stability, reactivity, and optical/electronic behavior. This approach offers deep insights into the functional consequences of doping beyond what static structural data alone can provide. Fig. 8 shows the corresponding maps for six difference combinations.

**Fig. 8.** Charge difference maps for six subtraction combinations. The S between brackets denote insertion of sulfur, whereas syn and anti-denote the corresponding symmetry.

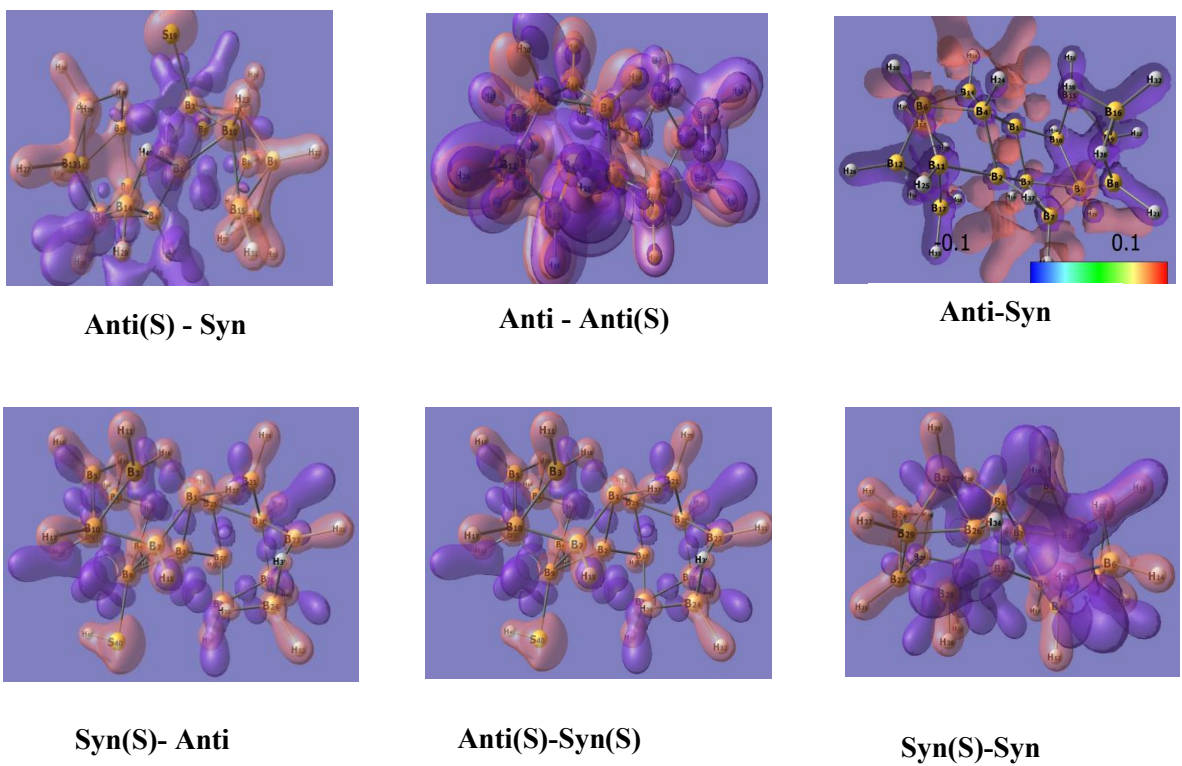

      Anti(S) - Syn        Anti - Anti(S)        Anti-Syn

      Syn(S)- Anti        Anti(S)-Syn(S)        Syn(S)-Syn

Theses diagrams reveal distinctive electronic charge redistributions between the syn- and anti-isomers of borane molecules, both in their pristine form and with sulfur substitution. These maps vividly highlight zones where electron density increases (red/orange) and decreases (purple/blue), showcasing how sulfur incorporation induces both localized and widespread electronic rearrangements.

This redistribution effect arises from sulfur's higher electronegativity and its ability to attract or repel electron density, thus creating active electronic regions centered on the sulfur atom and its neighboring sites. In contrast to the relatively uniform electronic distributions of unsubstituted molecules, where differences primarily stem from geometric variations between isomers, sulfur substitution triggers more profound electronic modifications.

These electronic properties changes include enhanced electrochemical activity through sulfur's redox versatility, increased electronic stability by stabilizing or attracting negative charge, particularly influential on sulfide ions, and the activation of molecular sites crucial for catalytic and binding functions.

In addition, sulfur improves charge mobility and increases the number of electronically active sites, leading to enhanced optical amplification efficiency and more stable laser emission. The sulfur-induced charge redistribution affects the spatial distribution and energies of molecular orbitals, thereby modulating optical absorption and emission characteristics vital for tailoring the performance of borane-based laser materials.

The 3-dim visualizations of molecular orbitals and charge differences in Fig. 8 further confirm that sulfur doping is a powerful approach to tailor the electronic and photophysical features of borane molecules, supporting their superior performance in photonics and advanced laser technologies. This comprehensive study positions sulfur-substituted boranes as promising candidates for next-generation optoelectronic and laser device materials.

### 4.6 NMR Calculations:

Nuclear Magnetic Resonance (NMR) spectroscopy is an indispensable tool for characterizing molecular structures. The synergy between experimental NMR and computational chemistry has grown significantly, with computational methods now capable of accurately predicting NMR parameters [32]. These calculations can aid in spectral assignment, structure refinement, and the study of dynamic processes. Presenting theoretical NMR calculations of chemical shifts of the borane (syn-, anti- unsubstituted & substituted) isomers under varying dispersion conditions helps to understand how dispersion interactions affect the electronic environment of the boron atoms influencing the NMR chemical shifts, and elucidating the role played by the substituting sulfur.

**figure 9.** NMR – chemical shift spectra for **a.** $B_{18}H_{22}$-syn and **b.** $B_{18}H_{22} - anti$, **c.** $1HS - B_{18}H_{21}$-syn and **d.** $4HS - B_{18}H_{21} - anti$

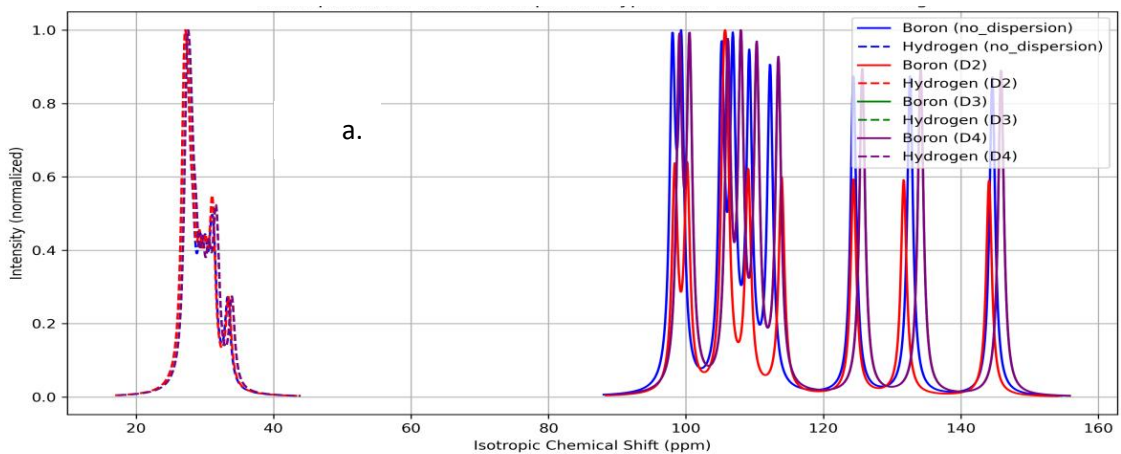

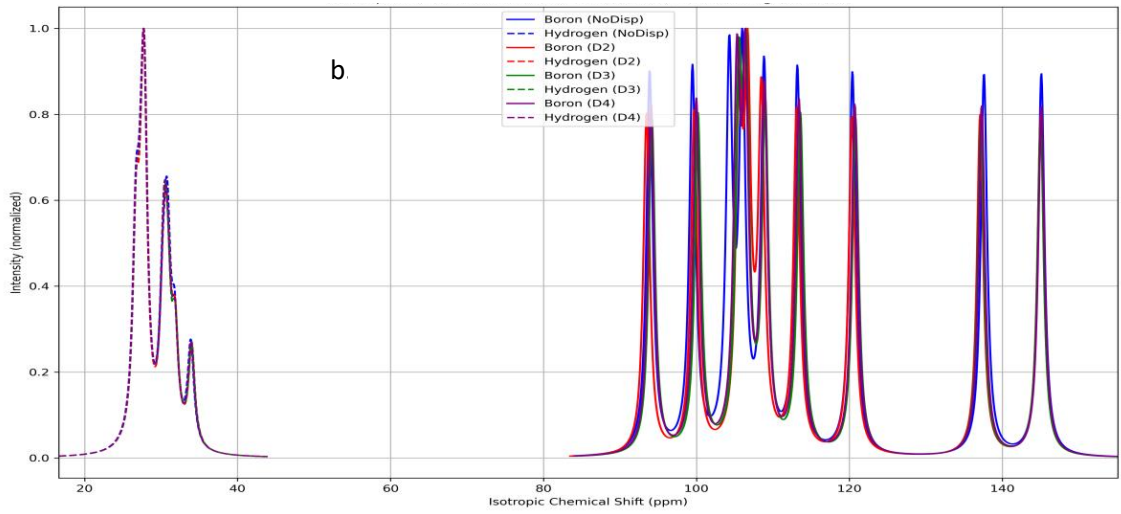

b.

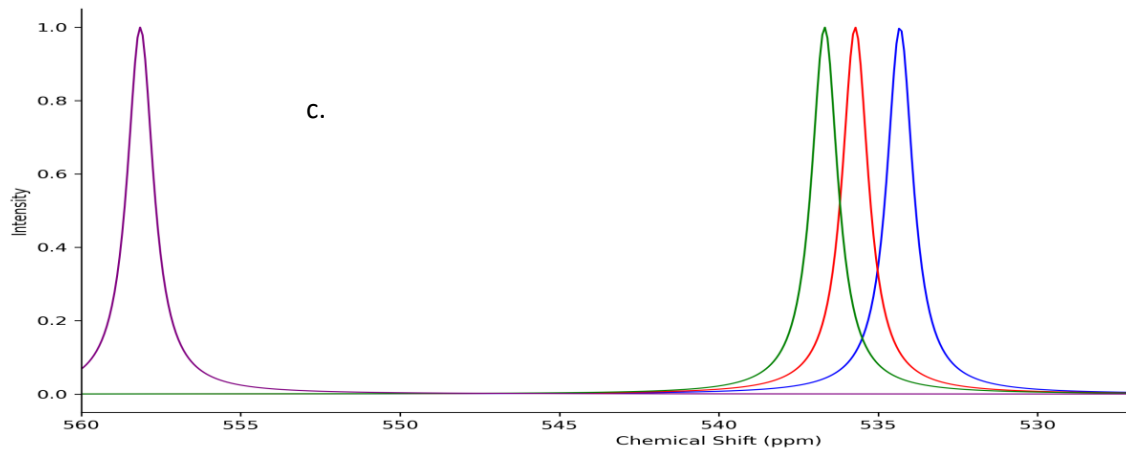

c.

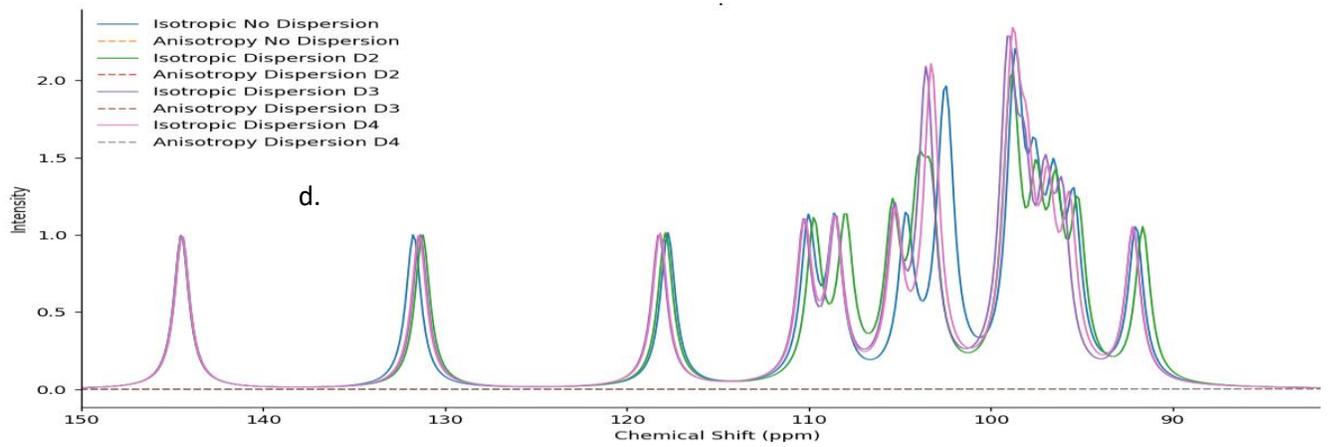

d.

The results showing the effect of the dispersion corrections and the role of substitution are shown in figure 9.

Figure (9.a) shows the NMR spectra of the syn- unsubstituted isomer, where a distinct peak is observed for both boron and hydrogen, reflecting a relatively high degree of molecular symmetry and rigidity. Also, a lower level of dispersion correction indicates a higher degree of rigidity in this particular configuration.

The results of (9.b) for the anti- unsubstituted borane, shows a similar set of peaks in the range of 90-150 ppm. This could be due to how the hydrogen molecules of the dipole change mass. Variations here may be influenced by changes in dipolar interactions and mass distribution of hydrogen atoms

For the syn- sulfur-substituted Borane, figure (9.c) shows that the dispersion corrections have a low impact. We see a sulfur peak cluster in the range of 530-560. Low dispersion has affected the data while the spectrum appears more complex than in the corresponding unsubstituted isomer.

Likewise, figure (9.d) shows the corresponding spectra for the anti-substituted isomer, where we see a regular distribution of peaks down to 100, with no clear effect of dispersion, followed by irregular peaks subsiding for values smaller than 90 ppm, with clear dispersion effects.

Actually, each isotope presents anisotropic and isotropic values, showing how the magnetic shielding changes as the electrons move through the molecule. Sulfur, being more electronegative than boron or hydrogen, introduces a significant perturbation to the electronic structure of the borane cage. This leads to changes in the chemical shifts of nearby nuclei, particularly those directly bonded to or in close proximity to the sulfur atom.

The emergence of a distinct sulfur resonance provides direct spectroscopic evidence of successful substitution. Discrepancies in spectra upon inserting sulfur may be due to variations in spatial interactions between the sulfur substituent and the borane cage.

Importantly, the spectra hinted that accurate modeling of non-covalent interactions via dispersion corrections is essential for reliable prediction of NMR parameters.

### 4.7. Electronic properties:

Borane clusters exhibit unique structural and electronic properties that make them attractive for various applications, including laser technology. We aim here to elucidate the effects of both the sulfur substitution and the dispersion corrections on the electronic properties (EHOMO, ELUMO, HLG, IP, EA, EF, $\eta$, $\omega$, and S) of selected borane isomers.

### 4.7.1. Functional and Dispersion Effects on Gap and Excitation Energies of Syn- and Anti- $B_{18}H_{22}$ Clusters with/without Sulfur Doping

- **The HOMO/LUMO and gap PB86/def2-SVP energies**

Density Functional Theory (DFT) has become a powerful tool for investigating the electronic structure and properties of borane molecules, which are known for their unique deltahedral cage geometries and rich chemistry. By employing the DEF2-SVP basis set, it is possible to accurately optimize molecular geometries and analyze electronic properties, including the nature and distribution of frontier molecular orbitals.

The Highest Occupied Molecular Orbital (HOMO) and the Lowest Unoccupied Molecular Orbital (LUMO) are particularly significant, as they play a central role in determining the chemical reactivity and stability of molecules [33]. Visualization and analysis of these orbitals provide insight into the electronic structure, potential reactive sites, and the overall electronic properties of borane species. HOMO and LUMO represent frontier orbitals critical for charge-transfer processes, optical properties, and redox behavior.

Table (10) shows the EHOMO values for $B_{18}H_{22}$, in both syn- and anti- forms, range from -0.3312 eV (D0) to -0.3287 eV (D4). The ELUMO values vary more significantly with dispersion corrections, ranging from -0.2249(-0.3219) eV to -0.2241(-0.2318) eV for the syn (anti)- isomer. In a similar way, the corresponding values of EHOMO (ELUMO) for the substituted 1HS-B18H18H21-syn range from -0.3329(-0.1906) V to -0.3226(-0.3293) V, with alike results for the 4HSB18H21-anti isomer.

**Table 10.** Electronic properties calculations of isomerism for the various Borane isomers, in equilibrium geometry, using the DFT method in the BP86/def2-SVP basis set.

| Electronic properties | B18H22-syn | | | | B18H22-anti | | | | 1HS-B18H21-syn | | | | 4HS-B18H21-anti | | | |
|---|---|---|---|---|---|---|---|---|---|---|---|---|---|---|---|---|
| | D0 | D2 | D3 | D4 | D0 | D2 | D3 | D4 | D0 | D2 | D3 | D4 | D0 | D2 | D3 | D4 |
| $E_{Homo}$ | -0.33 | -0.33 | -0.33 | -0.33 | -0.33 | -0.33 | -0.33 | -0.33 | -0.33 | -0.33 | -0.33 | -0.32 | -0.32 | -0.32 | -0.32 | -0.32 |
| $E_{Lumo}$ | -0.22 | -0.23 | -0.22 | -0.22 | -0.32 | -0.17 | -0.23 | -0.23 | -0.19 | -0.33 | -0.33 | -0.19 | -0.25 | -0.25 | -0.25 | -0.25 |
| HLG | 0.106 | 0.101 | 0.105 | 0.105 | 0.011 | 0.155 | 0.096 | 0.096 | 0.142 | 0.005 | 0.003 | 0.133 | 0.076 | 0.076 | 0.076 | 0.076 |
| IP | 0.333 | 0.334 | 0.341 | 0.345 | 0.187 | 0.326 | 0.327 | 0.33 | 0.301 | 0.313 | 0.308 | 0.311 | 0.196 | 0.306 | 0.307 | 0.31 |
| EA | 0.076 | 0.077 | 0.066 | 0.07 | 0.216 | 0.078 | 0.077 | 0.081 | 0.079 | 0.073 | 0.077 | 0.081 | 0.382 | 0.076 | 0.076 | 0.08 |
| $E_F$ | -0.28 | -0.28 | -0.28 | -0.28 | -0.33 | -0.25 | -0.28 | -0.28 | -0.26 | -0.33 | -0.33 | -0.26 | -0.28 | -0.28 | -0.28 | -0.28 |
| η | 0.053 | 0.05 | 0.052 | 0.052 | 0.005 | 0.077 | 0.048 | 0.048 | 0.071 | 0.002 | 0.002 | 0.067 | 0.038 | 0.038 | 0.038 | 0.038 |
| ω | 0.727 | 0.779 | 0.73 | 0.73 | 9.919 | 0.403 | 0.816 | 0.816 | 0.481 | 23.21 | 34.22 | 0.492 | 1.06 | 1.055 | 1.052 | 1.054 |
| S | 18.81 | 19.82 | 19.12 | 19.12 | 185.2 | 12.92 | 20.86 | 20.86 | 14.05 | 425.5 | 625 | 15.02 | 26.42 | 26.32 | 26.25 | 26.28 |

figure 10.a. HOMO for $B_{18}H_{22} - syn$   figure 10.b. LUMO for $B_{18}H_{22} - syn$   figure 10.c. HOMO/D2/ for $B_{18}H_{22} - syn$   figure 10.d. LUMO/D2/ for $B_{18}H_{22} - syn$

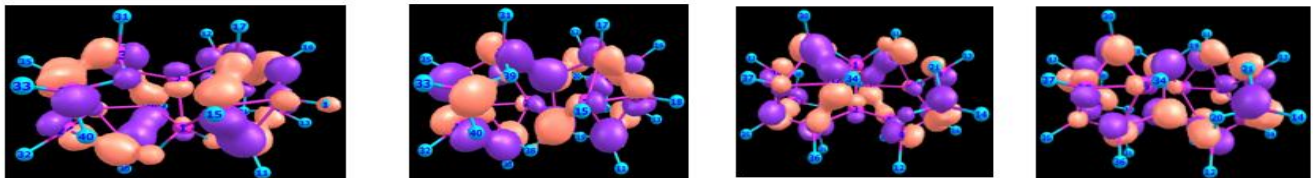

figure 10.e. HOMOD3/ for $B_{18}H_{22} - syn$   figure 10.f. LUMOD3/ for $B_{18}H_{22} - syn$   figure 10.g. HOMOD4/ for $B_{18}H_{22} - syn$   figure 10.h. LUMOD4/ for $B_{18}H_{22} - syn$

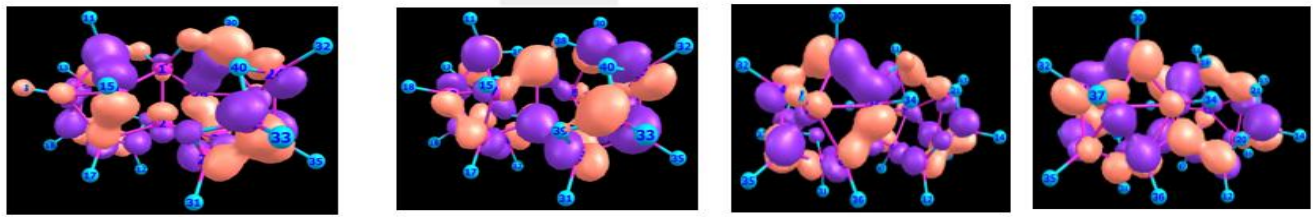

figure 11.a. HOMO/ for1HS $B_{18}H_{21} - syn$   figure 11.b. LUMO/ for1HS $B_{18}H_{21} - syn$   figure 11.c. HOMOD2/ for1HS $B_{18}H_{21} - syn$   figure 11.d. LUMOD2/ for1HS $B_{18}H_{21} - syn$

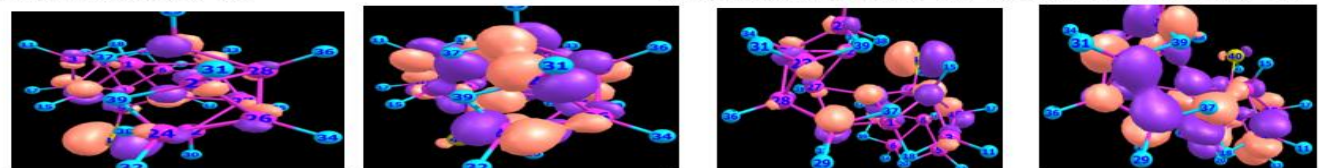

figure 11.e. HOMO D3/ for1HS $B_{18}H_{21} - syn$   figure 11.f. LUMO D3/ for1HS $B_{18}H_{21} - syn$   figure 11.g. HOMO D4/ for1HS $B_{18}H_{21} - syn$   figure 11.h. LUMO D4/ for1HS $B_{18}H_{21} - syn$

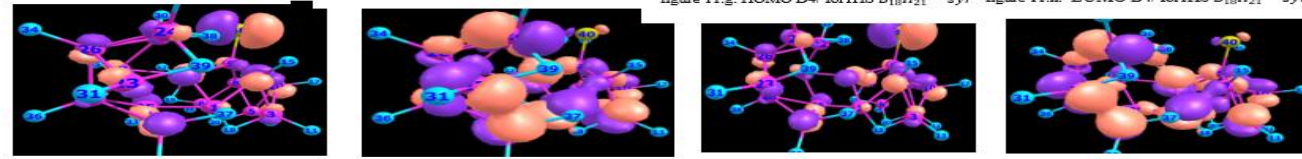

figure 12.a. HOMO / for $B_{18}H_{22} - anti$   figure 12.b. LUMO/ for1HS $B_{18}H_{22} - anti$   figure 12.c. HOMO D2/ for $B_{18}H_{22} - anti$   figure 12.d. LUMO D2/ for $B_{18}H_{22} - anti$

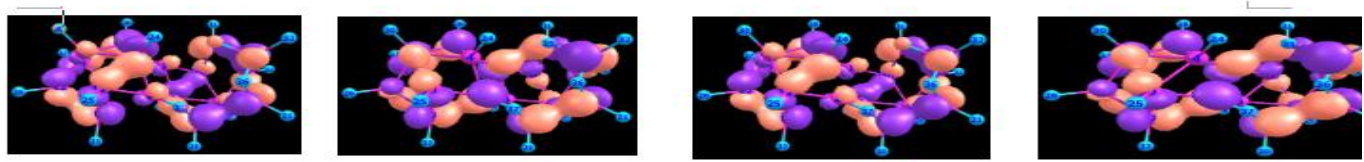

figure 12.e. HOMO D3 / for $B_{18}H_{22} - anti$   figure 12.f. LUMO D3/ for1HS $B_{18}H_{22} - anti$   figure 12.g. HOMO D4/ for $B_{18}H_{22} - anti$   figure 12.h. LUMO D4/ for $B_{18}H_{22} - anti$

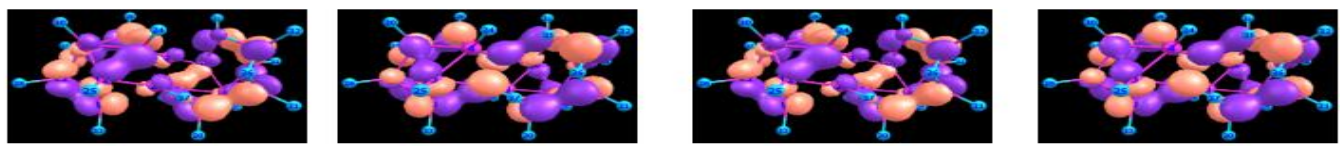

figure 13.a. HOMO / for4HS $B_{18}H_{21} - anti$   figure 13.b. LUMO/ for4HS $B_{18}H_{22} - anti$   figure 13.c. HOMO D2/ for4HS $B_{18}H_{21} - anti$   figure 13.d. LUMO D2/ for4HS $B_{18}H_{21} - anti$

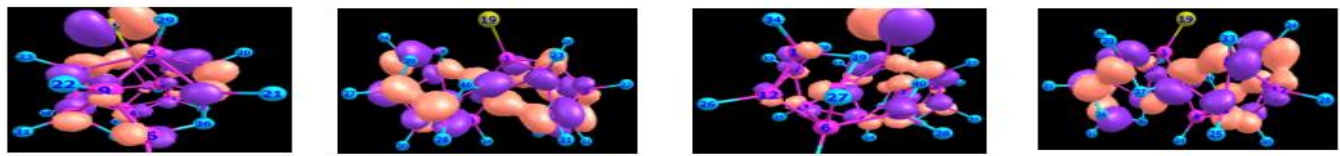

figure 13.e. HOMO D3 / for4HS $B_{18}H_{21} - anti$   figure 13.f. LUMO D3/ for4HS $B_{18}H_{21} - anti$   figure 13.g. HOMO D4/ for $B_{18}H_{21} - anti$   figure 13.h. LUMO D4/ for $B_{18}H_{22} - anti$

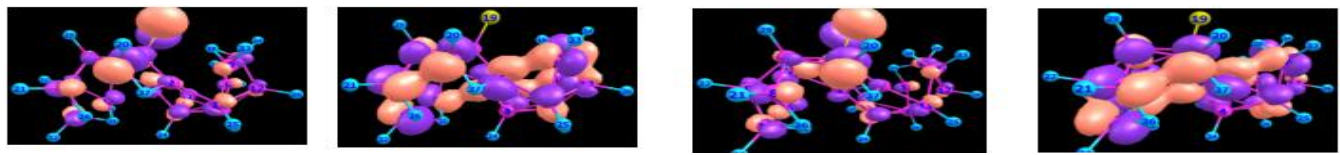

Figures 10 (11) give graphical representation, in the case of syn-unsubstituted (substituted) -and similarly for the anti- form in figures 12 (13)- of the HOMO (a) and LUMO (b) in the absence of dispersion corrections, and the corresponding ones with D2 (c,d), D3 (e,f) and D4 (g,h) corrections.

The HOMO in syn-$B_{18}H_{22}$ isomer, as shown in Fig. 10, exhibits a delocalized π-type character predominantly distributed over the central borane cage framework. The orbital demonstrates a pseudo-symmetrical distribution, featuring pronounced lobes localized around atoms 15, 33, and 40, which may represent reactive sites amenable to nucleophilic attack. The corresponding LUMO displays a complementary spatial distribution, with electron density concentrated near atoms 31 and 35.

Incorporation of dispersion corrections (D2, D3, and D4) induces subtle yet meaningful alterations in the frontier orbital distributions. Under the D2 correction, the HOMO exhibits enhanced localization around specific boron vertices, whereas the LUMO assumes a more diffuse character. The D3 correction further refines orbital topology, amplifying contributions from atoms 33 and 15. The most sophisticated D4 correction produces a more equilibrated electron density distribution across the cage, indicating that higher-order dispersion interactions play a critical role in stabilizing the electronic structure of these intricate borane clusters.

In Figure 11 for the 1HS-$B_{18}H_{21}$-syn isomer, the HOMO now incorporates substantial sulfur character, with electronic contributions extending to adjacent boron atoms, particularly atoms 31, 38, and 24, reflecting effective electronic communication between the substituent and the borane scaffold. Likewise, the LUMO features pronounced lobes around atoms 31 and 39, with notable extension toward the sulfur center.

Again, incorporating dispersion corrections exerts more pronounced effects on the sulfur-substituted derivative relative to the parent $B_{18}H_{22}$ cluster. The D2 correction prompts significant reorganization of both HOMO and LUMO distributions, whereas the D3 correction further localizes the HOMO around S–B bonding regions, while concomitantly promoting greater LUMO delocalization across the cage. The D4 correction yields the most balanced and physically consistent orbital distributions

In a similar way, the anti-$B_{18}H_{22}$ isomer, as depicted in Figure 12, exhibits frontier molecular orbital distributions that are notably distinct from those of the syn isomer. The HOMO demonstrates a more symmetrical electron density distribution, with significant contributions localized at boron atoms positioned at opposite vertices of the cage (notably atoms 25 and 34). Conversely, the LUMO displays a complementary pattern, with electron density concentrated primarily in the central region of the cage framework.

Dispersion corrections applied here induce trends similar to those observed in the syn conformer; however, the magnitude of these effects is somewhat attenuated, likely due to the more extended spatial arrangement in the anti-isomer that reduces intramolecular dispersion interactions.

Figure 13, for 4HS-$B_{18}H_{21}$-anti, engenders distinct electronic consequences relative to the 1-substituted syn- derivative. The HOMO features substantial sulfur orbital participation, with electron density extending toward adjacent boron atoms, particularly atoms 23 and 25, resulting in an asymmetric electron distribution across the cage. The LUMO similarly exhibits pronounced electron density near atoms 27 and 37, with notable sulfur involvement, especially evident in calculations incorporating D2 and D4 dispersion corrections.

In fact, dispersion corrections significantly influence the frontier orbitals of 4HS-$B_{18}H_{21}$-anti; in particular, D3 and D4 corrections reveal enhanced electronic communication between the sulfur center and remote boron atoms.

- **Functionals/Dispersion Efect on Gaps/Excitation Energy of Various Isemres**
  The effects of exchange-correlation functionals and dispersion corrections on the energy gap and excitation energy values for syn- and anti- $B_{18}H_{22}$ clusters, with and without sulfur doping, are illustrated in histograms 1 and 2 respectively (see appendix 1 for details).
  
  The histograms highlight the variations in electronic structure calculations, implemented with def2-SVP basis set, when changing the functionals, in order to account for the asymptotic exchange behavior at large distances, for all the four isomers with/without dispersion restricted to D4. They quantify how different computational methods influence the predicted energy gaps/excitation energies, revealing the subtle but significant impact of dispersion forces and sulfur substitution on the electronic properties. In addition to the functional PB86, used throughout the paper and which does not correct for large distance behavior, we consider in this subsection for comparison purposes two functionals, LC-WPBE and CAM-B3LYP, taking respectively 100% and 65 % asymptotic corrections.

**Histogram1:** Energy Gaps of various isomers using three functionals, LC-WBPE, CAM-B3LYP and BP86.

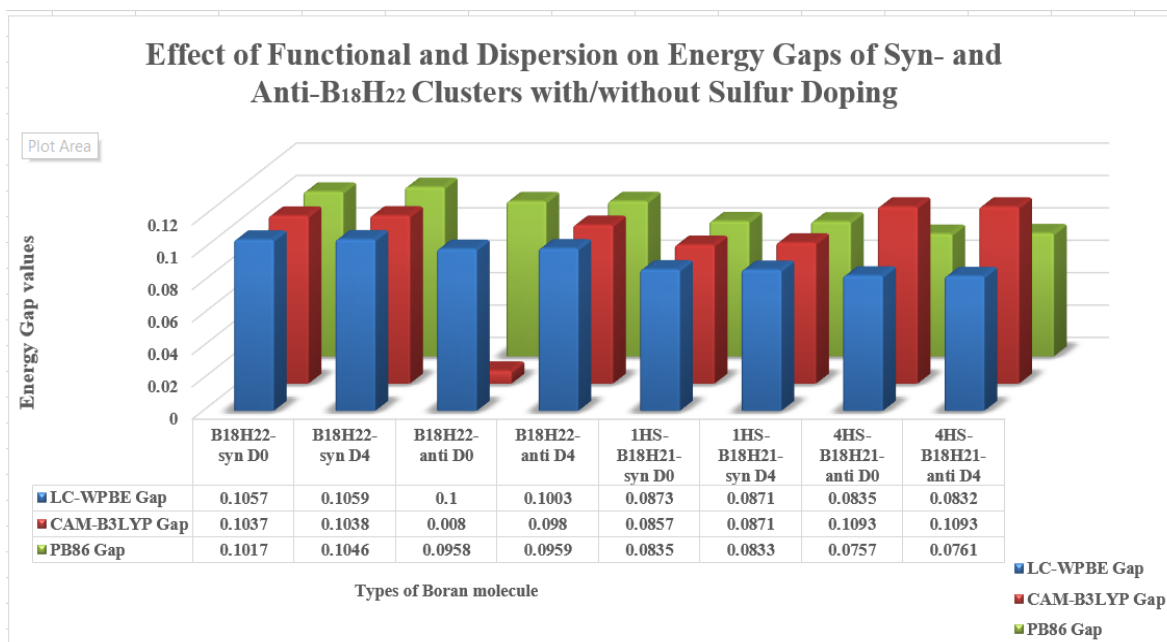

By comparing these clusters across isomeric forms and doped states, the histograms provide clear visual illustrations of how methodological choices and chemical modifications tune the furthest orbitals and related optical and electronic behaviors critical for potential laser and optoelectronic applications.

Relative percentage differences of the energy gap $E_g$, illustrated in Hist. 1 using the def2-SVP basis set, are presented in Table 11 for the three functionals. In order to estimate the functionals' influence on geometry changes, we state in the second column the relative energy gap difference ($\Delta E_g\%$) between the 'syn. (S)'-geometry and the 'anti (A)'-geometry for $B_{18}H_{22}$ for a fixed dispersion (D0 or D4). In contrast, we show in the third (fourth) column, estimating the functionals' influence on the dispersion, the relative energy gap difference ($\Delta E_g\%$) between the two dispersion states D0 and D4, for the fixed geometry syn. S (anti. A) of 1HS-$B_{18}H_{21}$ (4HS-$B_{18}H_{21}$).

It is clear that the variation in ($\Delta E_g\%$) due to dispersion interactions in $B_{18}H_{22}$ is minimal for the LC-WPBE functional, while it is noticeable for the PB86 method, and huge for the CAM-B3LYP method. The configurational effect, between A- and S- geometries, determining ($\Delta E_g\%$), deviates, in the absence of dispersion, much in CAM-B3LYP compared to PB86 and LC-WPBE, which yield both essentially equivalent values. When incorporating D4 dispersion, LC-WPBE and CAM-B3LYP show similar behaviors, while PB86 deviates by nearly 34%.

In the case of 1HS-$B_{18}H_{21}$ isomer, with S-geometry, evaluating ($\Delta E_g\%$) due to dispersion interactions, LC-WPBE and PB86 again provide consistent results, whereas CAM-B3LYP differs by approximately 87.5%. This trend slightly intensifies when considering 4HS-$B_{18}H_{21}$ with A-geometry, where both D0 and D4 give equal $E_g$ using CAM-B3LYP. In the absence of experimental data, it remains difficult to conclusively favor one functional over the others.

**Table 11.** Various combinations of $\Delta E_g\%$ for three functionals (LC-WPBE, CAM-B3LYP, PB86). A(S) denotes anti.(syn.)-geometry.

| $\Delta E_g\%$ | $B_{18}H_{22}\left(\frac{A-S}{S}\right)\%$ | | $1HS-B_{18}H_{21}\left(\frac{D_4-D_0}{D_0}\right)\%$, S. | $4HS-B_{18}H_{21}\left(\frac{D_4-D_0}{D_0}\right)\%$, A. |
|---|---|---|---|---|
| | $D_0$ | $D_4$ | | |
| **LC-WPBE** | 5.4 | 5.3 | 0.2 | 0.4 |
| **CAM-B3LYP** | 92.3 | 5.6 | 1.6 | 0 |
| **PB86** | 5.8 | 8.3 | 0.2 | 0.1 |

Moreover, sulfur substitution results in a notable reduction of the LUMO-HOMO energy gap ($Eg$), as Hist. 1 shows for all functionals. For instance, fixing the LC-WPBE functional, we see that $Eg$ decreases from around 0.1057–0.1059, according to dispersion corrections degree, in syn-unsubstituted boranes ($B_{18}H_{22}$) to roughly 0.0871–0.0873 upon sulfur substitution (1HS-$B_{18}H_{21}$) with same syn-geometry. This diminished gap facilitates easier electron excitation and transfer by lowering the photon energy threshold, which is beneficial for laser applications that require efficient optical response in specific spectral regions or under low-energy excitation.

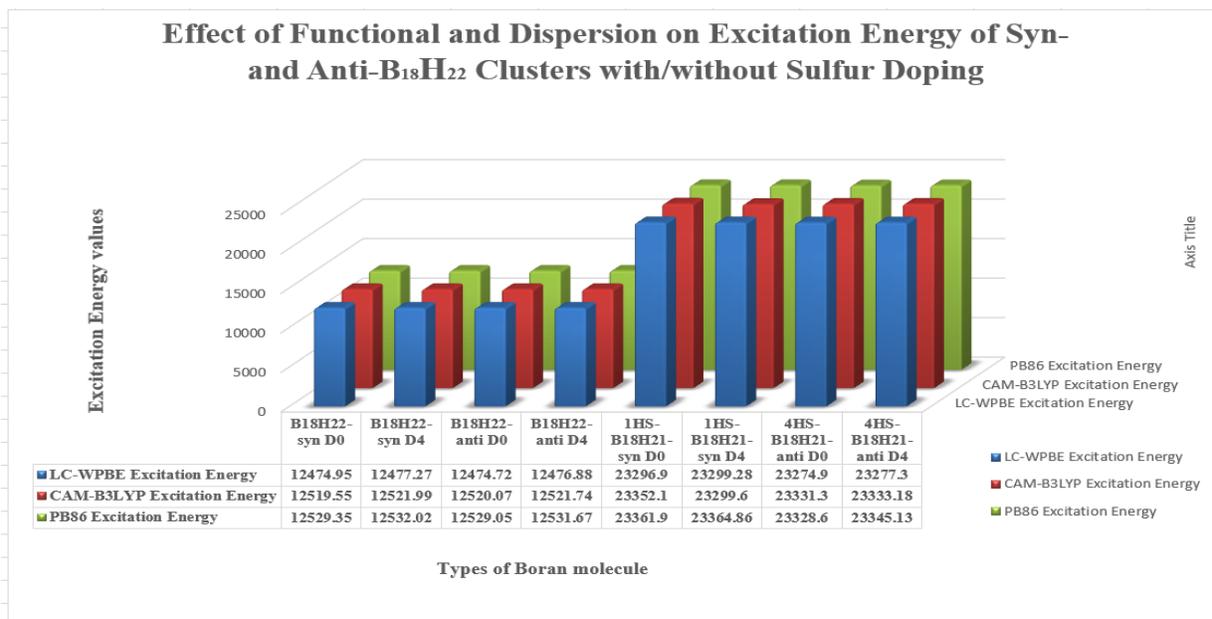

Histogram 2: Excitation Energy of various isomers using three functionals, LC-WBPE, CAM-B3LYP and BP86.

Similar to $E_g$ in Hist. 1, Histogram 2 corresponds to $E_{excitation}$. Unlike Hist. 1 where $E_g$ tends to diminish upon substitution, we see in Hist. 2 that $E_{excitation}$ tends to increase. Likewise, we state in table 12 the same relative energy deviations but corresponding to $E_{excitation}$. However, most of these relative deviations are minute compared to them in table 11.

**Table 12.** Various combinations of $\Delta E_g\%$ for three functionals (LC-WPBE, CAM-B3LYP, PB86). A(S) denotes anti.(syn.)-geometry.

| $\Delta E_{excitation}\%$ | $B_{18}H_{22}\left(\frac{A-S}{S}\right)\%$ | | $1HS-B_{18}H_{21}\left(\frac{D_4-D_0}{D_0}\right)\%$, S. | $4HS-B_{18}H_{21}\left(\frac{D_4-D_0}{D_0}\right)\%$, A. |
|---|---|---|---|---|
| | $D_0$ | $D_4$ | | |
| **LC-WPBE** | 0.0 | 0.0 | 0.0 | 0.0 |
| **CAM-B3LYP** | 0.0 | 0.0 | 0.2 | 0.0 |
| **PB86** | 0.0 | 0.0 | 0.0 | 0.1 |

Looking at both histograms and tables 11 and 12, one can conclude that the various functionals give approximately similar results, and so the theoretical model PB86-D4/def2-SVP can provide quantitatively reliable results for predicting new structures within the borane laser family with credible photophysical properties. This ability, joint with simplicity, offers valuable guidance to experimentalists for synthesizing, characterizing, and pumping new lasers based on borane derivatives.

### 4.7.2. Ionization Potential:

IP is defined as the energy required to remove an electron from the molecule, and is approximated given by the negative of the EHOMO energy. Table 10 states the IP values, as calculated through [34]:

$$IP = E_{cation} - E_{neutral}$$

While the IP values in $B_{18}H_{22}$ -syn range from 0.333 to 0.345 eV, the anti-isomer exhibits a lower IP of 0.187 eV without dispersion, increasing to around 0.326-0.330 eV with dispersion corrections.

The 1HS-$B_{18}H_{21}$-syn isomer shows an IP ranging from 0.300 to 0.313 eV with dispersion corrections. In contrast, the 4HS-$B_{18}H_{21}$-anti isomer has a lower IP of 0.196 eV without dispersion, increasing to 0.306-0.310 eV with dispersion.

When sulfur is directly bonded (as in 1HS and 4HS isomers), it drags an electron density from the borane cage, generally stabilizing the molecule. The inclusion of dispersion corrections (D2, D3, D4) generally leads to an increase in IP values, suggesting that dispersion forces stabilize the neutral species more than the cation.

### 4.7.3. Electronic affinity:

The electron affinity (EA) expresses the capability of the molecule to attract electrons. It is evaluated via [34]:

$$EA = E_{neutral} - E_{anion}$$

EA values for the syn isomer is around 0.076 eV and consistently lower than that of the anti-isomer, which give 0.216 eV without dispersion and range between 0.077-0.081 eV with dispersion.

The EA values for 1HS-$B_{18}H_{21}$-syn range from 0.073 to 0.081 eV whereas the corresponding range for 4HS-$B_{18}H_{21}$-anti isomer is considerably higher: 0.382 eV without dispersion, decreasing to approximately 0.076-0.080 eV with dispersion.

The sulfur atom's electron-dragging nature tends to increase the EA, as the resulting anion is stabilized by the presence of the electronegative sulfur. The position of the sulfur atom (1HS vs. 4HS) also plays a crucial role. Actually, the 4HS isomer shows a more pronounced effect on the EA, particularly without dispersion corrections whose effect generically reduces the EA. This means that dispersion forces stabilize the anion less than the neutral species.

### 4.7.4. Fermi Energy:

The Fermi energy ($E_F$) in the context of HOMO and LUMO orbitals for molecules is typically defined as the energy level at which the probability of electron occupancy is 50% at absolute zero. For molecules and intrinsic semiconductors, the Fermi energy (or Fermi level) is conventionally taken as the midpoint between the HOMO and LUMO energies [34]:

$$E_F = \frac{E_{Lumo} + E_{Homo}}{2}$$

The Fermi energy provides insight into the electrochemical potential of the electrons. The $E_F$ ranges from -0.278 (-0.33) eV to -0.276(-0.25) eV for $B_{18}H_{22}$ -syn(anti-), while for the substituted 1HS-$B_{18}H_{21}$-syn, the $E_F$ varies significantly with different dispersion corrections, ranging from -0.262 eV to -0.331 eV, whereas it is quasi constant (-0.28 eV) for 4HS-$B_{18}H_{21}$-anti, as shown in Table (10). Actually, sulfur substitution can lead to an electronic density displacement which affects the overall electronic properties of the molecule.

### 4.7.5. Hardness (η):

Hardness (η) quantifies a molecule's resistance to electron density changes. For boranes, dispersion corrections marginally alter η by modifying the IP and the EA. We adopt the definition [34]:

$$\eta = \frac{IE - EA}{2}$$

The hardness of $B_{18}H_{22}$ -syn (anti-) is about 0.053(0.01-0.05) eV. The substituted 1HS-$B_{18}H_{21}$-syn isomer shows significantly lower chemical hardness, especially with the D3 correction (0.0016 eV), indicating increased ductility and reactivity, whereas the corresponding value for the anti-conformer is quasi constant (0.04 eV), as shown in Table (10).

Dispersion corrections affect hardness and ductility through their effect on the HOMO-LUMO gap. Sulfur substitution dramatically changes the hardness and ductility, affecting the chemical stability and reactivity of the molecule.

### 4.7.6. Softness (S):

While chemical hardness (η) measures the resistance to charge transfer, the softness (S) refers to the molecule's ability to polarize, and can be defined as [34]:

$$S = \frac{1}{\eta}$$

### 4.7.7. Energy Gap:

The HOMO-LUMO gap ($E_g$ or HLG) is a fundamental quantum chemical property that provides insight into the molecule's electronic structure, chemical reactivity, and stability, and is given by [34]:

$$E_g = E_{Lumo} - E_{Homo}$$

Table 10 shows the relatively stable HOMO-LUMO gap for unsubstituted borane $B_{18}H_{22}$ across dispersion corrections, irrespective of its type (syn or anti). For the substituted 1HS-$B_{18}H_{21}$, the sulfur substitution significantly reduces $E_g$, especially with D2 and D3 corrections -indicating a red shift in absorption-, compared to the 4HS-$B_{18}H_{21}$-anti isomer, where we observe a similar $E_g$ to $B_{18}H_{22}$, suggesting that sulfur substitution at this position has less effect on electronic transitions.

### 4.7.8. Electrophilicity Index (ω):

The Electrophilicity Index (ω) is a quantum chemical descriptor that quantifies a molecule's ability to accept electrons and thus act as an electrophile, and is defined as [34]:

$$S = \frac{E_F^2}{2\eta}$$

From Table 10, we see that the **ω** of $B_{18}H_{22}$-syn is nearly constant, of about 0.73 eV, whereas for the 1HS-$B_{18}H_{21}$-syn isomer it shows a significantly increased value, especially with the D3 correction (34.2171 eV), indicating a strong ability to accept electrons.

### 4.7.9. Polarizability:

Polarizability is a fundamental property that describes the response of a molecule's electron cloud to external electric fields, underpinning a wide range of physical phenomena from dielectric behavior to optical activity. In molecular terms, polarization is quantified by the polarizability tensor, which measures the ease with which the electron distribution can be distorted, thereby inducing a dipole moment. This property is particularly significant in the context of borane hydride clusters, whose unique deltahedral architectures confer both structural rigidity and remarkable electronic flexibility. The anti-$B_{18}H_{22}$ cluster exhibits a highly symmetric, robust structure that supports a large, delocalized electron cloud, making it highly polarizable. This pronounced polarizability not only influences its fundamental optical properties, such as absorption and emission, but also enables advanced functionalities including efficient blue laser emission and photo-stability that surpass those of many organic dyes [35].

Table 13. Polarization calculations of isomerism for the borane various isomers in equilibrium geometry, using the DFT method in the BP86/def2-SVP basis set.

| Quantity | B18H22-syn | | | | B18H22-anti | | | | 1HS-B18H21-syn | | | | 4HS-B18H21-anti | | | |
|---|---|---|---|---|---|---|---|---|---|---|---|---|---|---|---|---|
| | D0 | D2 | D3 | D4 | D0 | D2 | D3 | D4 | D0 | D2 | D3 | D4 | D0 | D2 | D3 | D4 |
| $\alpha_{iso}$ (Å³) | 34.28 | 34.09 | 33.08 | 33.08 | 34.29 | 34.12 | 34.7 | 34.08 | 37.35 | 37.19 | 37.08 | 37.12 | 37.14 | 37.1 | 36.87 | 36.95 |

We limited our evaluation to the "isotropic" polarizability ($\alpha_{iso}$) defined as [36]:

$$\alpha_{iso} = \frac{1}{3}(\alpha_{xx} + \alpha_{yy} + \alpha_{zz})$$

From Table 13, we confer that the anti-$B_{18}H_{22}$ isomer consistently exhibits higher $\alpha_{iso}$ compared to the syn-$B_{18}H_{22}$ isomer across all dispersion corrections. This suggests that the former structure allows for a more easily perturbed electron cloud, likely due to its more delocalized electronic configuration and symmetry, than the latter structure.

The introduction of a sulfur substituent significantly enhances $\alpha_{iso}$ in both syn- and anti-isomers. This is attributed to the higher polarizability of sulfur compared to hydrogen, and the ability of sulfur to participate in electron delocalization within the borane cluster.

The application of dispersion corrections (D2, D3, and D4) generally leads to a reduction in the calculated $\alpha_{iso}$ values for all isomers. The magnitude of reduction varies depending on the specific dispersion model and the isomer's structure. This indicates that accounting for dispersion interactions modulates the electron density distribution, influencing the overall polarizability. However, the effect is smaller in the sulfur-substituted compounds.

The enhanced polarizability observed upon sulfur substitution suggests a potential pathway for improving the laser efficiency of borane clusters. A higher polarizability implies a greater capacity for induced dipole moments under the influence of an electromagnetic field, which can directly affect the molecule's ability to absorb and emit light efficiently.

## 5. Summary and Conclusion

This study presents a comparative analysis of the electronic properties of syn and anti $B_{18}H_{22}$ isomers and their sulfur-doped derivatives using DFT calculations.

It presents a novel exploration of sulfur doping's significant impact on the thermal stability and electronic properties of the borane clusters and their derivatives.

Substitution of boron atoms with sulfur leads to notable modifications in electronic structure, driven by localized charge redistribution and strengthened bonding, consistent with recent theoretical findings. The results of [37] demonstrated that sulfur inclusion in boron clusters induces a transition from linear chain to planar or quasi-planar configurations, accompanied by increased binding energies and enlarged HOMO-LUMO gaps, indicative of enhanced stability and electronic robustness.

Sulfur's role in promoting electron delocalization and reinforcing covalent interactions contributes to increased thermal resistance, a desirable attribute for optoelectronic materials. Supporting studies on chalcogen doping in boron-based nanomaterials further affirm sulfur's stabilizing effect. It was shown in [38] that sulfur doping enhances thermal stability and mechanical strength by fortifying atomic bonds and restricting atomic migration in semiconductor substrates. These results underscore sulfur's crucial contribution to both electronic and structural integrity in boron clusters, validating our approach for developing stable and tunable laser-active materials.

The results highlight the importance of dispersion corrections in accurately modeling the Borane systems and the effect of sulfur doping on enhancing cluster stability. The results indicate that sulfur-doped borane clusters are promising systems for optoelectronic applications, including tunable lasers, due to their modified electronic properties and enhanced stability. Further investigations, including excited-state calculations and planned experimental validation, are warranted to fully explore their potential in laser technology.

A critical component of our computational strategy is the comprehensive application of advanced dispersion corrections (D2, D3, and D4) to address the known limitations of traditional DFT in capturing weak van der Waals forces. The authors of [39] established the superior accuracy and physical grounding of the D4 method in predicting molecular structure and energy across diverse systems. Inclusion of these corrections mitigates errors in geometry predictions and enables precise modeling of photophysical properties sensitive to subtle structural variations.

Dispersion corrections play an especially pivotal role in sulfur-doped boron clusters, where the balance of covalent bonding and long-range interactions governs cluster stability and excited-state dynamics. Our results confirm that D2, D3, and D4 corrections refine optimized geometries and frontier orbital properties, enabling accurate reproduction of experimental observations and confident predictions vital for material design.

Our comprehensive analysis of frontier molecular orbitals in $B_{18}H_{22}$ isomers and their sulfur-substituted derivatives reveals the profound influence of both structural factors (conformation, substitution position) and computational methodology (dispersion corrections) on the electronic properties. The distinctive orbital distributions observed with sulfur substitution, particularly the enhanced involvement of the substituent in frontier orbitals, provide a structural basis for the altered electronic properties that make these compounds promising candidates for optoelectronic applications. Likewise, the progressive refinement of orbital descriptions with increasingly sophisticated dispersion corrections underscores the importance of accounting for these non-covalent interactions in accurate electronic structure calculations of borane clusters.

Previous studies on similar systems have used full-range-separated functionals with 100% asymptotic exchange to provide more accurate HOMO and LUMO levels (e.g. see [40–42]). As a test for the functional PB86 used in our work, we recalculated the excitation energy and energy gap with two other functionals, more sophisticated and better suited for taking into account the asymptotic behavior, however the results were similar to the PB86 ones, giving credence to our relatively simple but -arguably- powerful method.

The computational methodology combined good DFT/basis set to accurately predict geometries, IR/UV spectra, NMR chemical shifts, dipole moments, polarizability and excited-state properties, thereby offering a comprehensive theoretical understanding of syn, anti $B_{18}H_{22}$ and its sulfur doped isomers.

In summary, sulfur doping cooperatively improves thermal and electronic aspects of borane clusters, while state-of-the-art dispersion corrections are indispensable for correctly describing their complex structural and photophysical characteristics. This integrated methodology lays a strong foundation for advancing next-generation optoelectronic material design.

Our work broadly contributes to the deliberate design and electronic tuning of borane clusters for laser applications by correlating doping effects and structural evolution. Literature such as [43] highlights thermal stability challenges in chalcogen-hyperdoped silicon, illustrating doping's influence on electronic and optical traits, analogous to borane laser materials. [44] elucidates structural and aromaticity changes in sulfur-doped boron clusters that bolster chemical and electronic stability crucial to light emission, and [45] presents size-dependent stability and electronic structure characteristics in Bn and AlBn clusters pertinent to property modulation.

Collectively, these insights reveal that sulfur doping modulates electronic distributions and HOMO-LUMO gaps within borane clusters, enhancing photophysical responses like fluorescence and laser emission. The charge redistribution and structural reconfiguration induced by sulfur substitution optimize electronic stability, fluorescence performance, and laser gain, marking this doping strategy as key for engineering borane-based molecular laser materials with tunable optical properties and robust functionality for advanced photonics.

Our theoretical predictions, say of the enhanced thermal stability of sulfur-doped variants, and spectra can, in the future, be directly compared with newly experimentally available data (see [46–48] for newly acquired data).

To conclude, the findings can contribute significantly to the fundamental understanding of structure-property relationships in borane chemistry and provide valuable insights for the rational design of borane-based materials with specified electronic properties. We hope this work will raise awareness of the capabilities of substituted boranes, and, actually, using substitutes other than sulfur -as some of our current works in progress show- may open up channels for more sophisticated applications related to use of lasers.

**Funding:** This research did not receive any external funding.

**Data Availability Statement:** The data supporting the findings of this study are available within the article.

**Author Contributions:**

**Fakher Abbas**: Conceptualization, Methodology, Software, Writing- Original draft preparation, **Nabil Joudieh**: Conceptualization, Methodology, Writing- Original draft, Supervision, Project administrator, **Habib Abboud:** Supervision, **Nidal CHAMOUN:** Supervision, Validation, Writing- Reviewing and Editing,

**Conflicts of Interest:** The authors declare no conflict of interest.

**Acknowledgments:** N.C. acknowledges support from the CAS PIFI fellowship and the Alexander von Humboldt Foundation

# 6. Appendices

## Appendix 1

### Functionals & Basis Sets: Comparison, Strengths and limitations

In this appendix, we provide a concise yet comprehensive comparison of the PB86, LC-ωPBE, and CAM-B3LYP functionals, highlighting their fundamental characteristics and the critical roles played by long-range range-separated Hartree-Fock exchange and dispersion corrections in their performance. While our study tests the reliability of PB86 in the context of borane laser applications, this comparison underscores that range-separated hybrid functionals featuring full long-range Hartree-Fock exchange, such as LC-ωPBE combined with state-of-the-art dispersion corrections (D3/D4), offer significant advantages for accurately predicting electronic properties—including HOMO-LUMO gaps and excitation energies, in both sulfur-doped and undoped borane clusters. These benefits contrast with conventional generalized gradient approximations like PB86 and hybrid functionals with partial long-range exchange, such as CAM-B3LYP.

## 1) Motivations for Applying the PB86 Functional in this study

In this study, the PB86 functional was selected primarily for its well-established reliability in accurately optimizing geometries and describing ground-state properties of borane clusters, including those doped with sulfur. As a GGA functional, PB86 strikes an excellent balance between computational efficiency and accuracy, despite lacking exact Hartree-Fock (HF) exchange and the correct asymptotic behavior of the exchange potential.

While range-separated hybrid functionals such as LC-ωPBE and CAM-B3LYP improve long-range behavior by incorporating substantial amounts of exact HF exchange, 100% in LC-ωPBE and approximately 65% in CAM-B3LYP, resulting in more accurate frontier orbital energies and excitation energies, they come with a significantly higher computational cost. This added expense can be prohibitive for large cluster systems like those studied here. Additionally, tuning the wavelength of borane laser derivatives involves substituting hydrogen atoms with bulky substituents, which further increases computational demands.

Nonetheless, PB86 remains effective in producing qualitatively reliable charge difference maps and capturing essential electronic structure features relevant to doping effects and cluster stability. Its combination with empirical dispersion corrections enhances the modeling of noncovalent interactions and cluster geometries, making it a robust and practical choice for this system. Importantly, using PB86 also acts as a benchmark for assessing the functional's reliability in this application. It provides a foundation for future comparisons with more computationally intensive range-separated hybrids. Overall, PB86 enables manageable computational costs without sacrificing necessary accuracy for ground-state geometries, electronic trends, and charge density analysis, supporting the objectives of the present work. Where higher accuracy is required, selected excitation energies and gaps can be validated through calculations with LC-ωPBE or CAM-B3LYP, in line with best practices for computational studies of complex clusters.

## 2) Contrasted common functionals

- **PB86 functional**

    **Type:**

The PB86 functional, combining Becke's 1988 exchange with Perdew's 1986 correlation, is widely valued for its excellent balance of accuracy and computational efficiency, particularly in transition metal chemistry.

    **Advantages**:

    - PB86 is broadly implemented across quantum chemistry packages, ensuring accessibility.
    - It reliably predicts molecular geometries and bond lengths within 1 pm of experimental data for 3d metal complexes, outperforming other GGAs and methods like HF and

MP2, making it especially suitable for structural optimizations in inorganic and bioinorganic systems.
- As a pure GGA without exact exchange, PB86 enables faster calculations, facilitating studies of large molecules and extensive geometry optimizations.
- It benefits from systematic error cancellation, improving the reliability of relative properties such as bond length changes and energy differences, and provides robust transition-state geometries comparable to higher-level methods, supporting mechanistic investigations.

### Limitations:

- it suffers from self-interaction errors due to the absence of exact exchange, which can lead to overstabilization of delocalized states and less reliable results for charge-transfer and high-spin systems.
- It lacks explicit dispersion corrections, causing underestimation of binding energies in weakly interacting systems like van der Waals complexes and hydrogen bonds.
- Its performance in excited-state TD-DFT calculations is limited, especially for charge-transfer excitations and multiplet structures.
- Energetically, PB86 shows larger errors (5–10 kcal/mol) in reaction barriers and bond dissociation energies compared to hybrids like B3LYP (3–5 kcal/mol).
- While excellent for 3d metals, PB86 is less accurate for heavier 4d and 5d transition metals, where hybrid or meta-GGA functionals perform better.

### Dispersion:

The PB86 functional must be combined with empirical dispersion corrections (D2, D3, or D4) to account for noncovalent interactions accurately, as pure GGAs neglect London dispersion forces.

Overall, PB86's superior geometric accuracy, computational speed, and reliable performance with first-row transition metals make it a pragmatic choice for many computational chemistry applications, despite its energetic and excited-state limitations relative to hybrid functionals.

- **CAM-B3LYP Functional**
   ### Type:

   Range-separated hybrid functional combining the popular B3LYP functional with a long-range correction scheme using Coulomb-attenuating method (CAM).

   ### Advantages:

   - It incorporates a mixture of short-range and long-range Hartree-Fock exchange (approximately 19% short-range and 65% long-range exact exchange).

- It corrects the asymptotic decay of the exchange potential, improving the description of charge-transfer excitations and electronic spectra.
- It provides balanced accuracy for both ground and excited states.
- Well established and tested in many molecular systems with moderate computational cost.

**Limitations:**

The fraction of exact exchange is not 100% in the long-range limit (unlike some full range-separated hybrids), which may still lead to slight underestimation of excitation energies.

**Dispersion:**

Often used with empirical dispersion corrections like D3(BJ) for improved accuracy in noncovalent interactions.

- **LC-ωPBE Functional**

    **Type:**
    Full range-separated hybrid functional with 100% Hartree-Fock exchange in the long-range limit.

    **Advantages:**
    - The key feature is the explicit partitioning of exchange into short-range DFT and long-range 100% exact Hartree-Fock exchange, ensuring correct asymptotic behavior of the exchange potential.
    - It provides superior accuracy for frontier molecular orbital energies, band gaps, charge-transfer excitations, and excited-state properties.
    - It is especially effective for predicting HOMO-LUMO gaps and electronic spectra due to the full long-range HF exchange.
    - It often shows better consistency and systematic improvement over hybrid functionals without full long-range correction.

    **Limitations:**
    - It is computationally more demanding due to the full HF exchange at long range.
    - Short-range correlation might be less accurate than in some other hybrids.

    **Dispersion:**
    Requires addition of empirical dispersion corrections (D3 or D4) for proper modeling of van der Waals interactions, as dispersion is not well captured by pure DFT exchange-correlation functionals.

### 3) Importance of Long-Range Range-Separated Hartree-Fock Exchange

Delocalization error, a common flaw in approximate density functionals, leads to inaccurate electron density predictions and substantial quantitative errors. Modern computational approaches, including global and range-separated hybrid functionals, effectively reduce these errors and simultaneously handle related challenges such as dispersion interactions.

Pure generalized gradient approximations (GGAs) like PB86 notably fall short in reproducing the correct long-range decay of the exchange potential, resulting in significant deviations in orbital energies, particularly underestimating HOMO-LUMO gaps and excitation energies, especially for charge-transfer states.

Range-separated hybrids, exemplified by functionals like CAM-B3LYP and LC-ωPBE, overcome these limitations by partitioning the electron-electron interaction into short-range and long-range components. The short-range exchange is described by density functional theory approximations, while the long-range part is treated using exact Hartree-Fock exchange.

LC-ωPBE goes further by incorporating 100% Hartree-Fock exchange at long range, ensuring accurate asymptotic behavior and superior precision in frontier orbital energies and excitation states, as consistently demonstrated in the literature. CAM-B3LYP, in contrast, uses a significant but less than complete portion (approximately 65%) of long-range Hartree-Fock exchange, balancing computational cost with accuracy effectively.

This strategy of long-range exchange achieves a more reliable and physically sound description of excited state properties critical for accurate modeling of complex molecular systems.

### 4) Role of Dispersion Corrections (see appendix 2)

All these functionals require empirical dispersion corrections (D2, D3, or D4) because DFT's local and semi-local exchange-correlation approximations cannot capture noncovalent van der Waals interactions, which are crucial for accurately reproducing:
- Structural parameters and relative stability.
- Interaction energies.
- Vibrational and photophysical response.

D4 is the latest generation and improves over D2/D3 by including charge-dependent dispersion coefficients, leading to better accuracy and transferability across a wide range of systems. LC-ωPBE and CAM-B3LYP combined with D3 or D4 corrections yield highly reliable predictions in complex systems, including doped clusters and optoelectronic materials.

### 5) Basis Set Used

- **The def2-SVP Basis Set: Balancing Precision and Computational Efficiency**

The def2-SVP basis set is a split-valence polarized double-zeta basis set widely utilized in computational chemistry due to its excellent balance between accuracy and computational efficiency.

**Advantages**:

- It is particularly well-suited for organic and main-group chemistry applications, providing reliable results for geometry optimizations and electronic property calculations while maintaining a moderate computational cost.
- By including polarization functions on all atoms, including hydrogen, def2-SVP offers an improved description of electron density distortions and molecular geometries compared to non-polarized or partially polarized basis sets.
- As part of the systematic def2 family, it delivers consistent performance across the periodic table, with benchmark studies showing mean absolute deviations of approximately 8.9 kJ/mol for atomization energies at the Hartree-Fock level and around 2.0 kJ/mol with density functional theory methods.
- The basis set is optimized for modern computational techniques such as DFT, ensuring stable self-consistent field convergence and making it a common choice for initial geometry optimizations before progressing to larger basis sets like def2-TZVP or def2-QZVP for higher precision.
- Additionally, def2-SVP is frequently employed in hybrid basis set approaches, where it is applied to less critical atoms in order to reduce computational cost while maintaining accuracy for key atoms treated with larger basis sets.

**Limitations**:

- While def2-SVP offers a practical compromise between speed and accuracy, it may not be sufficient for high-precision energy calculations, where larger basis sets are preferable.

Overall, def2-SVP remains a valuable and versatile tool in computational chemistry, enabling efficient and reliable simulations across a wide range of molecular systems.

# Appendix 2

## Dispersion Corrections: Advances, Efficiency, and Applications

### 1) The Critical Role of Dispersion Corrections

London dispersion forces -weak attractive interactions arising from electron density fluctuations- are critical in computational chemistry, particularly in systems dominated by noncovalent bonding, such as molecular crystals, biomolecules, and supramolecular assemblies.

Conventional density functional theory (DFT) often fails to account for these forces, leading to inaccuracies in energy predictions, molecular geometries, and stability assessments. To address this, dispersion-corrected DFT methods (e.g., DFT-D2, DFT-D3, DFT-D4) introduce empirical post-self-consistent field (SCF) energy terms based on atomic positions, types, and, in advanced cases like D4, partial charges.

These corrections enhance accuracy without significant computational overhead, enabling reliable modeling of weak interactions in complex systems.

While methods such as many-body dispersion (MBD) or nonlocal functionals (e.g., VV10) offer detailed many-body or nonlocal treatments, they demand substantially more resources. In contrast, DFT-D3 and D4 balance efficiency and precision, excelling in large-scale studies of molecular stability, reactivity, and material design.

By bridging theoretical predictions with experimental observations, these corrections have become indispensable tools for advancing drug discovery, nanotechnology, and materials science, underscoring their transformative role in modern computational chemistry.

**2) Dispersion Corrections: Essential for Ground-State Computational Accuracy**

Dispersion forces play a vital role in computational chemistry by contributing significantly to the stabilization of molecular structures, especially in systems that are sterically crowded or possess large nonpolar surfaces.

A notable example is the stability of hex phenyl ethane derivatives, which classical bonding theories alone cannot explain but can be attributed to substantial London dispersion effects.

These forces are also key components of various noncovalent interactions such as π-π stacking, van der Waals complexes, and host-guest interactions within supramolecular chemistry. Such interactions are fundamental to processes like molecular recognition, aggregation, and self-assembly, which are crucial in fields like materials science and drug development.

Additionally, dispersion interactions are essential for accurately modeling surface and interface phenomena, such as the adsorption of molecules on materials like graphene. Neglecting dispersion in these contexts often leads to significant underestimation of binding energies and interaction characteristics.

Furthermore, the traditional perspective of steric hindrance as purely repulsive has evolved, recognizing that dispersion forces can counterbalance repulsive effects, thereby stabilizing structures and enabling reactivity patterns that might otherwise appear unfavorable.

**3) Impact of Dispersion on Excited-State Lifetimes:**

Dispersion interactions critically shape excited-state dynamics by modulating molecular packing, structural preferences, and energy landscapes.

In systems like diphenyl ether, dispersion stabilizes specific conformations or aggregated states in the excited state, directly influencing nonradiative decay pathways and lifetimes. For example, ultrafast spectroscopy and computational studies reveal that dispersion-driven structural arrangements can either extend or shorten excited-state durations by altering relaxation efficiency.

In solid-state systems, such as metal-free phthalocyanine films, dispersion governs chromophore proximity and packing density, which dictate exciton lifetimes by controlling energy transfer rates and nonradiative decay channels. Tighter packing from stronger dispersion can enhance quenching or stabilize excitons, depending on the system's electronic structure.

Dispersion also contributes to the interaction energies of excited-state complexes (e.g., excimers or exciplexes), modifying potential energy surfaces and altering barriers for processes like internal conversion or intersystem crossing. These adjustments directly impact dissociation rates and lifetimes.

Additionally, in photochemical systems like molecular switches, dispersion stabilizes key transition states or intermediates during photoisomerization, modulating the efficiency and speed of excited-state transitions. By balancing stabilization and steric effects, dispersion fine-tunes the lifetimes of photoactive states, underscoring its pivotal role in designing optoelectronic materials and light-driven molecular machines.

## 4) Dn Dispersion Corrections (n=2,3,4): Methodology, Many-Body Effects, and Parameterization

To address the limitations of standard DFT in modeling dispersion, Stefan Grimme and collaborators developed a series of empirical dispersion correction methods; D2, D3, and D4; that are now widely used in computational chemistry

- **The D2 Dispersion Correction Method: Foundations and Limitations:**

Introduced in 2006, the D2 method pioneered the integration of pairwise atom-atom dispersion interactions into density functional theory (DFT) calculations. This approach approximates dispersion energy as a sum of two-body terms, utilizing fixed, element-specific coefficients and a damping function to prevent overcounting short-range electron correlation effects.

While D2 significantly enhanced the modeling of noncovalent interactions (e.g., van der Waals forces), its simplicity comes with trade-offs: the coefficients remain static, unaffected by an atom's chemical environment or coordination state, leading to inaccuracies in systems with diverse bonding scenarios.

The method's parameters -predetermined dispersion coefficients and a universal scaling factor tailored to the DFT functional- ensure computational efficiency and ease of implementation. However, the lack of environmental adaptation limits its reliability for complex materials or dynamically changing molecular systems.

Despite these shortcomings, D2 remains a low-cost, accessible tool for preliminary studies, though it is often superseded by more advanced corrections (e.g., D3, D4) in modern computational workflows requiring higher precision.

- **D3 Dispersion Correction: Improved Accuracy via Environmental Adaptation:**

Introduced in 2010, the D3 method addressed key limitations of its predecessor (D2) by incorporating environment-dependent dispersion coefficients, enabling dynamic adjustments based on an atom's local bonding environment (e.g., coordination number, oxidation state).

This advancement significantly improved accuracy across diverse systems, from small molecules to extended materials. Unlike D2's fixed coefficients, D3 adapts dispersion interactions to reflect chemical context, better capturing subtle variations in noncovalent forces.

Building on pairwise atom-atom interactions, the D3 method introduces two pivotal advancements to refine dispersion modeling. First, it optionally incorporates three-body dispersion effects via the Axilrod–Teller–Muto term, which captures collective interactions among atom triplets. This addition improves accuracy in densely packed or extended systems where many-body electron correlations are significant. Second, D3 employs adjustable damping schemes (e.g., zero-damping or Becke–Johnson damping) to modulate short-range interactions, ensuring seamless integration with the underlying density functional theory (DFT) framework. These damping functions prevent overcounting electron correlation at close distances while preserving the accuracy of long-range dispersion forces. Together, these innovations enhance D3's adaptability across diverse chemical environments, though the optional three-body terms may incur higher computational costs.

While D3's three-body terms enhance precision, they increase computational cost, leading many implementations to exclude them by default.

Despite this trade-off, D3 remains widely adopted in quantum chemistry software due to its balance of accuracy and efficiency. Its environmental sensitivity makes it particularly effective for studying heterogeneous materials, supramolecular systems, and surfaces, cementing its role as a cornerstone of modern dispersion-corrected DFT.

- **D4 Dispersion Correction: Breakthroughs and Broad-Spectrum Applications**

The D4 dispersion correction, introduced in 2019, represents a significant advancement in modeling noncovalent interactions within density functional theory (DFT) and related computational methods.

Unlike earlier corrections, D4 dynamically adjusts dispersion coefficients using atomic partial charges, making the treatment of dispersion sensitive to both atom type and its electronic environment. This approach enables a more accurate, transferable, and robust modeling of dispersion forces, especially in large, complex, or charged systems.

D4 also incorporates a more sophisticated and consistent treatment of many-body dispersion effects, further improving accuracy for highly polarizable or charged species.

The coefficients in D4 are calculated on-the-fly using Casimir-Polder integration and electronic density information, which enhances its adaptability and reliability across diverse chemical environments.

In terms of performance, D4 consistently delivers high accuracy for a wide range of systems, including molecular crystals, inorganic salts, and surface adsorption phenomena.

It outperforms or matches both its predecessor D3(BJ) and more computationally demanding methods, particularly in predicting solid-state polarizabilities and adsorption energies.

D4's environment-dependent parameterization and many-body effects provide a physically improved model that is applicable to both molecular and periodic systems, including those involving heavy and highly coordinated elements.

Despite its advanced features, D4 remains computationally efficient, adding only a modest overhead compared to D2 or D3. This efficiency, combined with its accuracy, makes it suitable for routine DFT calculations as well as for use with semi-empirical and force-field methods, supporting high-throughput and large-scale simulations.

D4's robustness is further demonstrated by its successful application to large molecular systems, such as supramolecular complexes and biomolecular assemblies, where it maintains reliable performance across different chemical environments.

However, D4 is not without challenges. Its complex parameterization requires careful reference data, especially for periodic and highly coordinated systems.

In rare cases, D4 can introduce unphysical features in the potential energy surface, such as artificial minima or bumps, though these are less frequent than in D3 and are being addressed through ongoing refinements like D4S and D4SL.

Additionally, while D4 shows major improvements in some properties, the gains over D3(BJ) for lattice energies and cell volumes of molecular crystals are less pronounced, as D3(BJ) already performs well in these areas.

Overall, D4 stands out for its environment-sensitive dispersion coefficients, improved treatment of many-body effects, and computational efficiency. It is now considered the state-of-the-art empirical dispersion correction, offering superior accuracy and transferability for a broad range of chemical systems, from small molecules to large biomolecular assemblies and periodic materials. Its ongoing development and refinement continue to enhance its reliability and applicability in modern computational chemistry.

In summary, the D4 correction bridges accuracy and efficiency, making it a cornerstone tool for modern computational studies-from supramolecular chemistry to actinide protein assemblies. Its ongoing development continues to expand its applicability in both academic and industrial research.

## 5) Alternative Dispersion Correction Approaches: Methods, Advantages, and Practicalities

Beyond the widely used D2, D3, and D4 corrections, several other approaches are available for incorporating dispersion interactions in computational chemistry, each with unique strengths and considerations.

Nonlocal correlation functionals, such as VV10, embed dispersion effects directly into the exchange-correlation functional by introducing a nonlocal correlation term that depends on electron densities at different points in space. This allows for a seamless, parameter-free treatment of long-range van der Waals forces, though it can be more computationally demanding than empirical corrections.

The Exchange-Dipole Model (XDM) takes a physically motivated approach, deriving dispersion coefficients from the electron density and its derivatives, making the correction adaptable to various chemical environments. However, its implementation is more complex and computationally intensive compared to simpler empirical methods.

The Tkatchenko-Scheffler (TS-vdW) model improves upon fixed-coefficient schemes by scaling atomic dispersion coefficients according to the effective atomic volume, as determined by the electron density. This makes the correction environment-sensitive and more accurate, albeit with some added computational overhead.

The Many-Body Dispersion (MBD) method goes even further by explicitly accounting for collective many-body effects using a model of coupled quantum harmonic oscillators, which is especially important for large or condensed-phase systems but comes at a higher computational cost.

Dispersion-Correcting Potentials (DCPs) offer another route, using atom-centered potentials to correct for missing dispersion interactions, which can be tailored for specific atoms and basis sets but require careful parameterization and may lack broad transferability.

Among these methods, VV10 stands out for its ability to directly integrate dispersion into the DFT framework, providing accurate geometries and binding energies, especially when paired with suitable exchange functionals.

In practice, DFT-D3 is often recommended for structure optimizations, with single-point VV10 calculations used to refine interaction energies. The DFT-Dn series remains popular due to its computational efficiency, broad applicability, and ease of integration with existing quantum chemistry software. While methods like MBD, XDM, and nonlocal functionals can offer higher accuracy for certain systems, they are generally more computationally intensive.

For most practical applications, DFT-Dn corrections strike an optimal balance between speed and improved accuracy, making them suitable for large-scale and high-throughput studies. Benchmarking studies confirm that D3 and D4 provide accuracy on par with leading alternatives for most neutral molecular systems, although all methods face challenges with ionic complexes and the choice of exchange-correlation functional remains a critical factor in overall performance.